\documentclass[12pt,preprint]{aastex}
\begin{document}

%\shorttitle{The updated calibration}
%\shortauthors{Pilyugin \& Thuan}

\title{The redshift evolution of oxygen and nitrogen abundances 
in emission-line SDSS galaxies}

\author{Trinh X. Thuan}
\affil{Astronomy Department, University of Virginia, 
P.O. Box 400325, Charlottesville, VA 22904-4325} 
\email{txt@virginia.edu}
\and
\author{Leonid S. Pilyugin}
\affil{ Main Astronomical Observatory
                 of National Academy of Sciences of Ukraine,
                 27 Zabolotnogo str., 03680 Kiev, Ukraine}
\email{pilyugin@mao.kiev.ua}
\and
\author{Igor A. Zinchenko}
\affil{ Main Astronomical Observatory
                 of National Academy of Sciences of Ukraine,
                 27 Zabolotnogo str., 03680 Kiev, Ukraine}
\email{zinchenko@mao.kiev.ua}

\begin{abstract}
The oxygen and nitrogen abundance evolutions with redshift 
and galaxy stellar mass in emission-line 
galaxies from the Sloan Digital Sky Survey (SDSS) are 
investigated. This is the first such study for nitrogen abundances, and it
 provides an additional constraint for the study of the chemical evolution of galaxies.
We have devised a criterion to recognize and exclude from consideration
active galactic nuclei (AGNs) and star-forming galaxies with 
large errors in the line flux measurements. To select star-forming galaxies 
with accurate line fluxes measurements, we require that, for each galaxy, 
the nitrogen abundances derived with 
various calibrations based on different emission lines agree. 
Using this selection criterion, subsamples of star-forming SDSS galaxies have been 
extracted from catalogs of the MPA/JHU group.   
We found that the galaxies of highest masses, those with masses $\ga$ 10$^{11.2}$M$_\sun$, 
have not been enriched in both oxygen and 
nitrogen over the last $\sim$ 3 Gyr: they have formed 
their stars in the so distant past that these 
have returned their nucleosynthesis products to the interstellar medium before $z$=0.25. 
The galaxies in the mass range from  $\sim$ 10$^{11.0}$M$_\sun$  
to  $\sim$ 10$^{11.2}$M$_\sun$ do not show an appreciable 
enrichment in oxygen, but do show some enrichment in nitrogen: 
they also formed their stars before $z$=0.25 but later 
in comparison to the galaxies of highest masses; 
these stars have not returned nitrogen to the 
interstellar medium before $z$=0.25 because they have not had enough 
time to evolve. 
This suggests that stars with lifetimes of 2--3 Gyr, in the 1.5--2 M$_\sun$ mass range, 
%and not in the 3--8 M$_\sun$ mass range as predicted by current stellar evolutionary models of intermediate-mass stars, 
contribute to the nitrogen production.
Finally, galaxies  with masses  $\la$ 10$^{11}$M$_\sun$ show 
enrichment in both oxygen and nitrogen during the last 3 Gyr:  they have undergone appreciable star formation 
and have converted  
up to $\sim$20\% of their mass into stars 
over this period. Both oxygen and nitrogen enrichments increase  
with decreasing galaxy stellar 
mass in the mass range from $\sim$10$^{11}$M$_{\sun}$ to $\sim$10$^{10}$M$_{\sun}$, 
then slightly decrease with further decrease of galaxy mass.   
\end{abstract}


\section{INTRODUCTION}

The study of various correlations between the global properties of 
galaxies is very important for the understanding of 
 their formation and evolution.
Among these correlations, the one between 
galaxy mass and metallicity is one of 
the most important. \citet{lequeuxetal1979} first showed that 
the oxygen abundance correlates well 
with total mass for a sample of dwarf irregular and blue compact 
galaxies, in the 
sense that the higher the total mass, the higher the heavy element content. 
Since the galaxy mass is a poorly known parameter, 
the luminosity -- metallicity
relation has often been considered instead of the mass -- metallicity 
relation. 
\citet{garnettshields1987} found that abundances of spiral galaxies  
correlate well with their luminosities. 
The luminosity -- metallicity relation for nearby galaxies has been considered
subsequently  
by many authors  
\citep[][among others]{garnettshields1987, skillmanetal1989,vilacostas1992,richermccall1995,
pilyugin2001b,zaritskyetal1994,pilyuginetal2004,gusevaetal2009}.  

In recent years, the number of available spectra of emission-line galaxies 
has increased dramatically due to the completion of several 
large spectral surveys. 
Measurements of emission lines in those spectra have been carried out 
for abundance determinations and investigation of the luminosity -- 
metallicity relation. 
Thus, \citet{melbournesalzer2002} have considered the luminosity -- 
metallicity relation for 519 galaxies in the KPNO International Spectroscopic 
Survey (KISS), \citet{lamareilleetal2004} for 6,387 galaxies in the 
2dF Galaxy Survey, \citet{tremontietal2004} 
for about 53,000 galaxies in the Sloan Digital Sky Survey (SDSS), 
and \citet{asarietal2007}
for 82,302 SDSS galaxies.

It has been found, however, that, when mass can be determined, 
the mass-metallicity correlation is considerably 
tighter than the luminosity--metallicity correlation, suggesting 
that mass may be a more meaningful physical parameter than luminosity.  
In the last decade, the evolution of 
the mass-metallicity relation with redshift has been examined by 
many investigators
\citep{lillyetal2003,savaglioetal2005,erbetal2006,cowiebarger2008,
maiolinoetal2008, lamareilleetal2009,laralopezetal2009}.  
In those investigations, oxygen abundances have been derived using  
various methods. 
The general conclusion from those studies is that the oxygen 
abundance change of star-forming galaxies over the last half of the age of the 
universe appears to be somewhat moderate, with 
$\Delta$(log(O/H)) $\sim$ 0.3   
or lower. This change is comparable to or less than
the scatter in the observed oxygen abundances  
(see Fig.10 of \citet{lillyetal2003}, Fig.13 of \citet{savaglioetal2005}, 
and Fig.6 of \citet{laralopezetal2009}).

Until now, no attention has been paid to the redshift evolution of nitrogen 
abundances in galaxies, despite the fact that they present several advantages 
for the study of the chemical evolution of galaxies. 
First, since at 12+log(O/H) $\ga$ 8.3, 
secondary nitrogen becomes dominant and the nitrogen abundance increases at 
a faster rate than the oxygen abundance \citep{henryetal2000}, 
then the change in nitrogen 
abundances with redshift should show a larger amplitude in comparison to 
oxygen abundances and, as a consequence, should be easier to detect. 
Furthermore, there is a time delay in the nitrogen production as compared 
to oxygen production \citep{maeder1992,vandenhoek1997,pagel1997}. 
This provides an additional constraint 
on the chemical evolution of galaxies. 
These reasons have led us 
to consider here not only  
the redshift evolution of oxygen abundances but also that of 
nitrogen abundances. To carry out such an investigation, it is necessary 
to derive accurate oxygen and nitrogen abundances.

The emission line properties of photoionized H\,{\sc ii} regions are governed 
by its heavy element content and by the electron temperature distribution 
within the photoionized nebula. 
In turn, the latter is controlled by the ionizing star cluster spectral 
energy distribution and by the chemical composition of the H\,{\sc ii} region. 
The evolution of a giant extragalactic H\,{\sc ii} region associated with a  
star cluster is thus 
caused by a gradual change in time of the integrated stellar 
energy distribution due to stellar evolution. This has been 
the subject of numerous 
investigations
\citep[][among others]{stasinska1978,stasinska1980,mccalletal1985,dopitaevans1986,
moyetal2001,stasinskaizotov2003,dopitaetal2006}. 
The general conclusion from those studies is that H\,{\sc ii} regions 
ionized by star clusters form a well-defined fundamental sequence in 
different emission-line diagnostic diagrams. 
The existence of such a fundamental sequence forms the basis of 
various investigations of extragalactic H\,{\sc ii} regions.

First, \citet{baldwinetal1981} have suggested that the position of an 
object in 
some well-chosen 
emission-line diagrams can be used to separate H\,{\sc ii} regions 
ionized by star clusters from other types of emission-line objects. 
This idea has found general acceptance and use. Thus, the 
[OIII]$\lambda$5007/H${\beta}$ vs [NII]$\lambda$6584/H${\alpha}$ 
diagram has been used widely to distinguish between  
star-forming galaxies and active galactic nuclei (AGNs).
In particular, the SDSS emission-line galaxies occupy a 
well-defined region shaped like  
the wings of a seagull  
\citep{stasinskaetal2008}. The left wing consists 
of star-forming galaxies while the right wing is attributed to AGNs. 
However, the exact location of the dividing line between H\,{\sc ii} regions 
and AGNs is still controversial  
\citep{kewleyetal2001,kauffmannetal2003,stasinskaetal2006}.

Second, \citet{pageletal1979} and \citet{alloinetal1979} have suggested that the positions 
of H\,{\sc ii} regions in some emission-line diagrams can be calibrated in 
terms of their oxygen abundances. This approach to abundance determination 
in H\,{\sc ii} regions, usually referred to as the ``strong line method'' has 
been widely adopted, especially in 
cases where the temperature-sensitive [O {\sc iii}] $\lambda$4363 line is not detected. 
Numerous relations have been suggested to convert 
metallicity-sensitive emission-line ratios into metallicity or 
temperature estimates 
\citep[][among many others]{
dopitaevans1986,
zaritskyetal1994, 
pilyugin2000,
pilyugin2001a,
pilyuginthuan2005,
pettinipagel2004,
tremontietal2004,
liangetal2006,
stasinska2006}

It should be noted that the classic T$_{\rm e}$ method, which relies 
on the electron temperature  T$_{\rm e}$ determined from the 
[O {\sc iii}] $\lambda$4363 line,
is also based
on the existence of the fundamental sequence of  H\,{\sc ii} regions. 
Indeed, when only one electron temperature, 
t$_3$ or t$_2$, is measured (t$_3$ and t$_2$ being 
the electron temperatures in the [O {\sc iii}] and [O {\sc ii}] zones, 
respectively)  
it is standard practice to use a t$_2$ -- t$_3$ relation to estimate 
the other temperature.
This relation is usually established on the 
basis of H\,{\sc ii} regions models which belong to the same 
fundamental sequence. 

Our study will be based on the SDSS data base of a million spectra. 
To study the redshift evolution of oxygen and nitrogen abundances, 
accurate oxygen and nitrogen abundance determinations are mandatory.
The determination of accurate abundances in H\,{\sc ii} regions
from SDSS spectra poses two problems. 
First, line fluxes in SDSS spectra are 
measured by an automatic procedure. This inevitably introduces large flux
errors for some objects. We need to devise 
a way to recognize those objects and exclude them from consideration. 
Second, the SDSS galaxy spectra span a large range of redshifts. 
There is thus an aperture-redshift effect in 
SDSS spectra. Indeed, they are obtained with 3-arcsec-diameter fibers. 
At a redshift of $z$=0.05 
the projected aperture diameter is $\sim$ 3 kpc, while it is  $\sim$ 15 kpc
at a redshift of $z$=0.25. 
This means that, at large redshifts, 
 SDSS spectra are closer to global spectra of 
whole galaxies, i.e. to that of  
composite nebulae including multiple star clusters, 
rather than to 
spectra of individual H\,{\sc ii} regions.   
One should then expect that some SDSS objects will not follow the fundamental 
H\,{\sc ii} region sequence. 
 These need also to be identified and excluded from our sample. 

We propose here a method to recognize objects that suffer 
from one or both of these problems. It is based on the idea  
that if {\it i)} an object belongs to the fundamental H\,{\sc ii} region 
sequence,
and {\it ii)} its line fluxes are measured accurately, then the different 
relations between the line fluxes and the physical characteristic of 
H\,{\sc ii} regions, 
based on different emission lines, should yield similar  
physical characteristics (such as electron temperatures and 
abundances) of that object. 

The relations used for the determination of oxygen and nitrogen 
abundances and for the selection of star-forming galaxies 
with accurate line flux measurements are derived in Section 2. 
The selected subsamples of SDSS galaxies are described in Section 3. 
The relations between redshift, galaxy mass and 
metallicity are discussed in Section 4.
Section 5 presents the conclusions.

Throughout the paper, we will be using the following notations for the line 
fluxes: \\ 
R$_2$ = [O\,{\sc ii}]$\lambda$3727+$\lambda$3729  
      = $I_{[OII] \lambda 3727+ \lambda 3729} /I_{{\rm H}\beta }$,  \\
N$_2$ = [N\,{\sc ii}]$\lambda$6548+$\lambda$6584  
      = $I_{[NII] \lambda 6548+ \lambda 6584} /I_{{\rm H}\beta }$,  \\
R$_3$ = [O\,{\sc iii}]$\lambda$4959+$\lambda$5007        = 
$I_{{\rm [OIII]} \lambda 4959+ \lambda 5007} /I_{{\rm H}\beta }$,  \\
The electron temperatures will be given in units of 10$^4$K, and 
the stellar masses of galaxies in solar units.

\section{METALLICITY AND TEMPERATURE CALIBRATIONS}

We noted above that many relations have been proposed to convert 
various metallicity-sensitive emission-line combinations into 
metallicity or temperature 
estimates. The oxygen abundances derived with these 
different calibrations are not 
in good agreement, with differences amounting 
up to 0.7 dex \citep{kewleyellison2008}. 
In fact, there appears to exist as many different oxygen abundance scales  
as there are calibrations.  
Which metallicity scale is the correct one? 
There is strong evidence in favour of 
the metallicity scale defined by the classic T$_{e}$ method, 
considered the most reliable 
one \citep[see the discussion in][]{pilyugin2003,bresolinetal2009b}. 
Indeed, the oxygen abundances derived with the T$_{e}$ method have been  
confirmed by high-precision model-independent determinations
 of the interstellar 
oxygen abundance in the solar vicinity, using high-resolution observations 
of the weak interstellar O\,{\sc i}$\lambda$1356 absorption line 
towards stars, 
and by recent determinations of stellar abundances. 
To put our derived abundances on the T$_e$ scale, we will use  
H\,{\sc ii} regions  
with T$_{e}$-based abundances as the basis of our calibrations.

To establish the relations between different line fluxes and the physical 
characteristic of H\,{\sc ii} regions, we will be using the 
data for high-metallicity H\,{\sc ii} regions  
in nearby galaxies compiled by \citet{pilyuginetal2009}. 
We have added to this compilation the recent spectrophotometric 
measurements of  \citet{savianeetal2008,bresolinetal2009a,bresolinetal2009b,estebanetal2009}. 
The oxygen and nitrogen abundances in those H\,{\sc ii} regions were rederived 
in an uniform way within the framework of the standard  H\,{\sc ii} region model with 
a two-zone temperature distribution within the nebula. The equations for the 
electron temperatures and oxygen ionic abundances were taken from 
\citet{izotovetal2006}.
The relation between the electron temperature t$_2$ 
within the O$^+$ zone and t$_3$,  
that within the O$^{++}$ zone, is 
\begin{equation}
t_2 = 0.672 t_3 + 0.314 .
\label{equation:t2t3}   
\end{equation}
This relation derived in \citet{pilyuginetal2009} is very close to the 
commonly used one \citep{campbelletal1986,garnett1992}.

The abundance of the nitrogen ion N$^+$ is derived from the equation  
\begin{equation}
\log\frac{N^+}{O^+} = \log\frac{N_2}{R_2}  + 0.273 - \frac{0.726}{t_2} - 
0.02 \log t_2 + 0.007 t_2 .
\label{equation:eno}   
\end{equation}
The total nitrogen abundance is determined from 
\begin{equation}
\log\frac{N}{H} = \log\frac{O}{H}  +  \log\frac{N}{O}
\label{equation:enh}   
\end{equation}
with the assumption
\begin{equation}
\frac{N^+}{O^+} = \frac{N}{O}  .
\label{equation:xx}   
\end{equation}

It is common practice, in constructing the calibration, to establish 
relations between the oxygen abundances and the strong-line fluxes. 
As a result, oxygen abundances are the best studied. In particular, 
the redshift evolution of oxygen abundances has been the subject of 
several recent investigations, as noted above.
Here, we will adopt a different method. 
We will calibrate the positions of H\,{\sc ii} regions in three 
different diagnostic diagrams in terms of nitrogen abundances, 
to which less attention has been paid. The oxygen abundance 
will then be estimated from the obtained nitrogen abundance and the N/O ratio using 
equations (\ref{equation:eno}) and (\ref{equation:enh}). The independent determination of 
nitrogen abundances and of their redshift evolution will provide an independent check of the conclusions derived from the study of the 
redshift evolution of oxygen abundances.

\begin{figure}
\begin{center}
\resizebox{0.66\hsize}{!}{\includegraphics[angle=0]{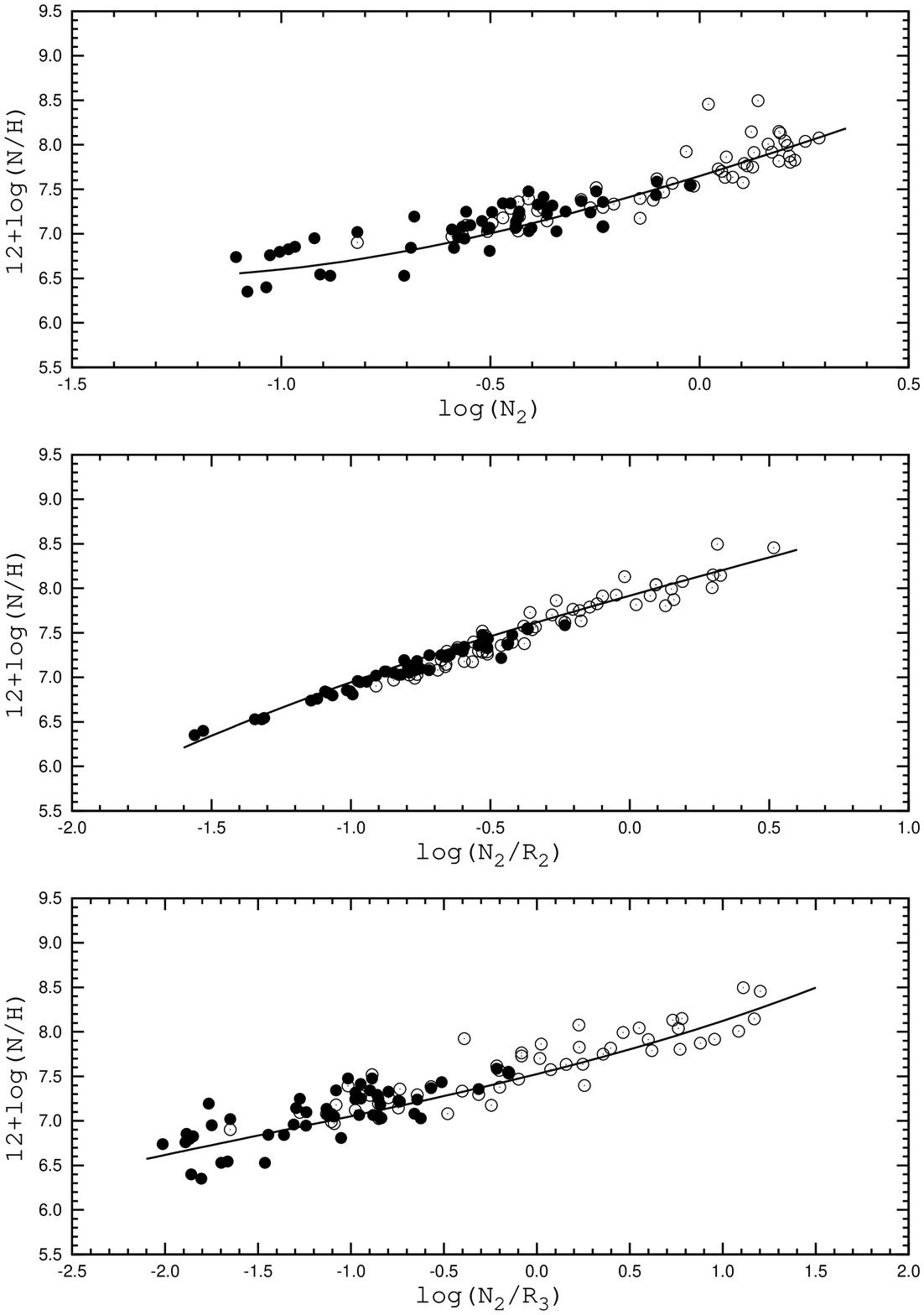}}
\caption{ 
Nitrogen abundance as a function of different abundance-sensitive indexes 
for H\,{\sc ii} regions in nearby galaxies with measured electron temperatures
t$_{3,O}$ or t$_{2,N}$. 
The filled circles show H\,{\sc ii} regions with measured t$_{3,O}$ temperatures.
The open circles show H\,{\sc ii} regions with measured t$_{2,N}$ temperatures.
The solid lines are the adopted cubic fits.
}
\label{figure:nh}
\end{center}
\end{figure}

Fig.\ref{figure:nh} shows the nitrogen abundance as a function of 
log(N$_2$) (upper panel), log(N$_2$/R$_2$) (middle panel), and 
log(N$_2$/R$_3$) (lower panel). 
The filled circles are H\,{\sc ii} regions with measured 
electron temperatures t$_{3,O}$ (the auroral line [OIII]$\lambda$4363 
is available). 
The open circles are H\,{\sc ii} regions with measured 
electron temperatures t$_{2,N}$ (the auroral line [NII]$\lambda$5755 
is available).
Those data have been fitted by the cubic expression 
\begin{equation}
12+\log(N/H) = a_0 + a_1X + a_2X^2 +a_3X^3
\label{equation:nh}   
\end{equation}
where X is successively  X=log(N$_2$), X=log(N$_2$/R$_2$), and X=log(N$_2$/R$_3$). 
The derived curves are shown in Fig.\ref{figure:nh} by solid lines.
The constant coefficients in Eq.(\ref{equation:nh}) for all three cases are given in  Table \ref{table:nht}.

\begin{figure}
\begin{center}
\resizebox{0.70\hsize}{!}{\includegraphics[angle=0]{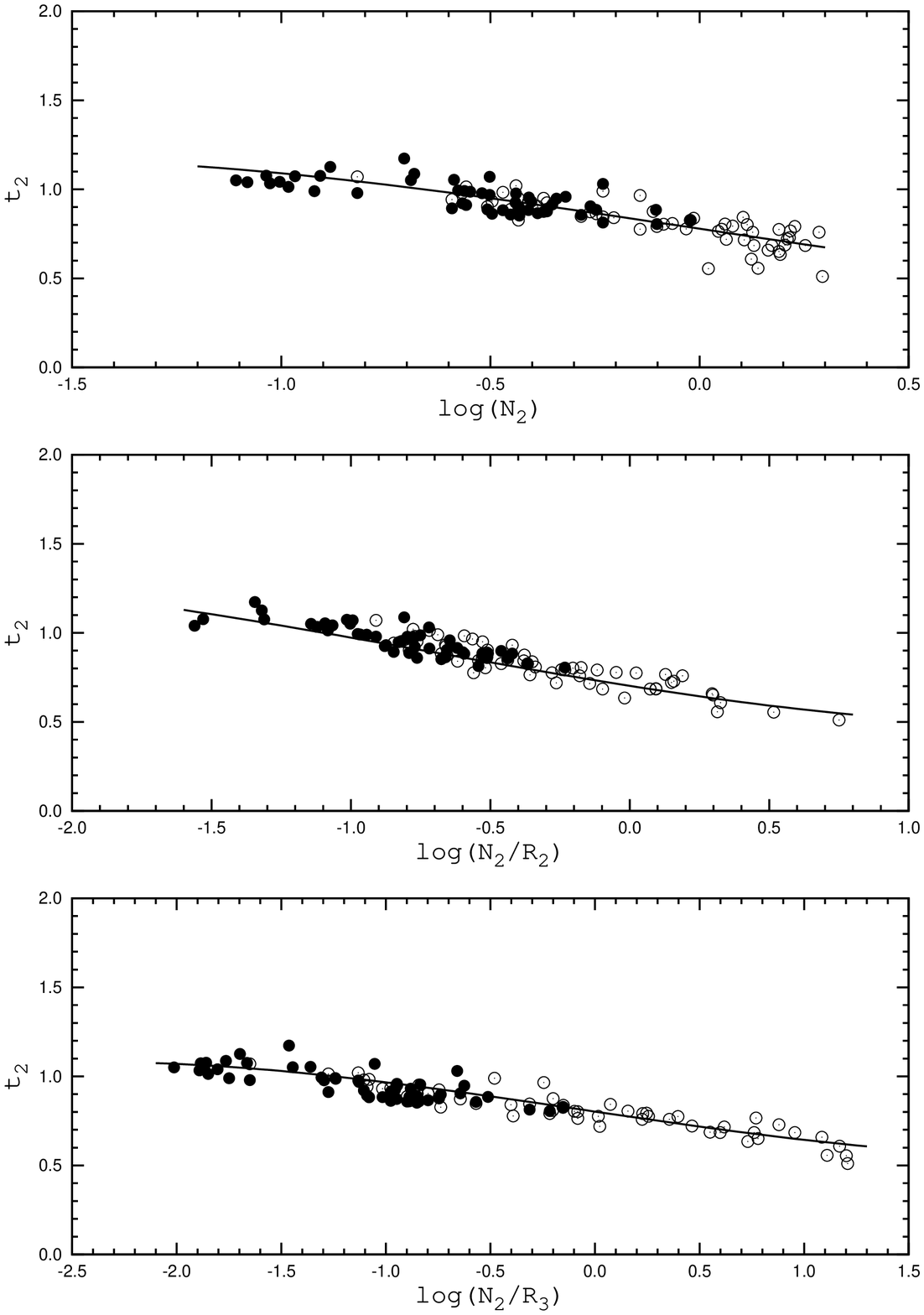}}
\caption{ 
Electron temperature t$_2$ as a function of different abundance-sensitive indexes 
for H\,{\sc ii} regions in nearby galaxies with measured electron temperatures
t$_{3,O}$ or t$_{2,N}$. 
The filled circles show H\,{\sc ii} regions with measured t$_{3,O}$ temperatures.
The open circles show H\,{\sc ii} regions with measured t$_{2,N}$ temperatures.
The solid lines are the adopted cubic fits.
}
\label{figure:t2}
\end{center}
\end{figure}

Fig.\ref{figure:t2} shows the electron temperature t$_2$ as a function of 
log(N$_2$) (upper panel), log(N$_2$/R$_2$) (middle panel), and 
log(N$_2$/R$_3$) (lower panel) for the same sample of H\,{\sc ii} regions.
Those data have also been fitted by the cubic expression 
\begin{equation}
t_2 = a_0 + a_1X + a_2X^2 +a_3X^3
\label{equation:t2}   
\end{equation}
where X is successively X=log(N$_2$), X=log(N$_2$/R$_2$), and X=log(N$_2$/R$_3$). 
The derived curves are shown in Fig.\ref{figure:t2} by solid lines.
The constant coefficients in Eq.(\ref{equation:t2}) for all three cases are  also given in  Table \ref{table:nht}. 

The relations between nitrogen abundances and the abundance-sensitive indexes 
N$_2$, N$_2$/R$_2$, and N$_2$/R$_3$ 
will be used for the determination of nitrogen abundances in 
the SDSS galaxies. They will also serve to select star-forming galaxies with 
accurate line fluxes measurements. 
As for the relations between the electron temperature t$_2$ and the 
N$_2$, N$_2$/R$_2$, and N$_2$/R$_3$ indexes, they will be used 
for estimating the N/O ratios 
and deriving the O/H oxygen abundances from the N/H nitrogen abundances.

\section{SAMPLE SELECTION}

Line flux measurements in SDSS spectra have been carried out by several 
groups. We use here the data in several catalogs made available publicly 
by the MPA/JHU group \footnote{The catalogs are available at 
http://www.mpa-garching.mpg.de/SDSS/}. These catalogs give   
line flux measurements, redshifts and various other derived physical properties  
such as stellar masses for a large sample of SDSS galaxies.
The techniques used to construct the catalogues are described 
in \citet{brinchmannetal2004,tremontietal2004} 
and other publications of those authors. We have chosen to use these 
catalogues instead of the original SDSS spectral database because they 
contain generally more accurate line flux measurements (see a discussion 
of the errors in the line flux measurements by \citet{brinchmannetal2004}).

In a first step, we extract from the MPA/JHU catalogs all emission-line 
objects with measured fluxes in the 
H$\beta$, H$\alpha$, [OII]$\lambda \lambda$3727,3729, [OIII]$\lambda$4959, 
[OIII]$\lambda$5007, [NII]$\lambda$6548, [NII]$\lambda$6584, [SII]$\lambda$6717, 
and [SII]$\lambda$6731 emission lines.  The hydrogen, oxygen and nitrogen lines 
will serve to estimate oxygen and nitrogen abundances relative to hydrogen, 
and the ratio of the sulfur line intensities is  an indicator of electron density. 
Since our calibrations are only valid in the low-density limit, we have included 
only those objects with a reasonable value of the [SII] ratio, i.e. those with 
1.25 $<$ F$_{[SII]\lambda 6717}$/F$_{[SII]\lambda 6731}$ $<$ 1.5. This results in 
a sample containing 118,544 objects which will be referred to hereafter 
as the total sample.
The wavelength range of the SDSS spectra is 3800 -- 9300 \AA\ so that
for nearby galaxies with redshift z $\la$ 0.02, the 
[O\,{\sc ii}]$\lambda$3727+$\lambda$3729 emission line is out of that range. 
The absence of this line prevents the determination of the 
oxygen abundance, so all SDSS galaxies with  z $\la$ 0.02 were also excluded. 
Thus, all galaxies in our total sample have redshifts greater than   
$\sim$ 0.023, i.e. they are more distant than $\sim$ 100 Mpc.
The redshift $z$ and stellar mass M$_S$ of each galaxy were also taken from 
the MPA/JHU catalogs.

The emission-line fluxes are then corrected for interstellar 
reddening using the theoretical H$\alpha$ to H$\beta$ ratio and the 
analytical approximation to the Whitford  interstellar reddening law 
from \citet{izotovetal1994}. In several cases, the derived 
value of the extinction c(H$\beta$) is negative and has been set to zero.

For each galaxy, we have estimated three values of the nitrogen abundance and three values 
of the electron temperature t$_2$, using the calibrations discussed 
above and the dereddened N$_2$, R$_2$, R$_3$ line intensities.
Since measurements of the [NII]$\lambda$6584 line are more reliable than those of  
the [NII]$\lambda$6548 line, we have used N$_2$ = 1.33[NII]$\lambda$6584 
instead of the standard 
N$_2$ = [NII]$\lambda$6548 + [NII]$\lambda$6584.  
Then the mean value N/H of the nitrogen abundance for each galaxy is determined as 
\begin{equation}
\log(N/H)_{mean} = \frac{\log(N/H)_{N_2} + \log(N/H)_{N_2/R_2}+ \log(N/H)_{N_2/R_3}}{3}. 
\label{equation:nhmean}   
\end{equation}
The mean value of the electron temperature t$_2$ is determined in a similar way. Using 
N/H and the mean  t$_2$, the N/O ratio and the oxygen abundance O/H are then derived 
using equations (\ref{equation:eno}) and (\ref{equation:enh}).

We have suggested above that if  {\it i)} the object belongs to the fundamental sequence 
of  HII regions,
and {\it ii)} its line fluxes are measured accurately, then the calibrations 
based on different lines should result in similar abundances. 
We have calculated for each galaxy the deviations of individual values of the nitrogen 
abundance from the mean value
$\Delta$log(N/H)$_{N_2, N_2/R_2, N_2/R_3}$ = 
log(N/H)$_{N_2, N_2/R_2, N_2/R_3}$ -- log(N/H)$_{mean}$. 
The mean deviation $\Delta$log(N/H)$_{mean}$ 
and the maximum deviation $\Delta$log(N/H)$_{max}$ were also computed for each object. 
We have then used the value of 
$\Delta$log(N/H)$_{max}$ as a selection criterion to extract 
from the total sample three subsamples of star-forming 
galaxies with accurate line flux measurements, going from the most stringent requirement to the least 
stringent one : \\
-- Subsample A contains only objects with $\Delta$log(N/H)$_{max}$ $\le$ 0.05, a total of 15,548 galaxies. \\ 
-- Subsample B contains only objects with $\Delta$log(N/H)$_{max}$ $\le$ 0.10, a total of 55,189 galaxies. \\ 
-- Subsample C contains only objects with $\Delta$log(N/H)$_{max}$ $\le$ 0.15, a total of 84,364 galaxies.

\begin{figure}
\begin{center}
\resizebox{1.00\hsize}{!}{\includegraphics[angle=0]{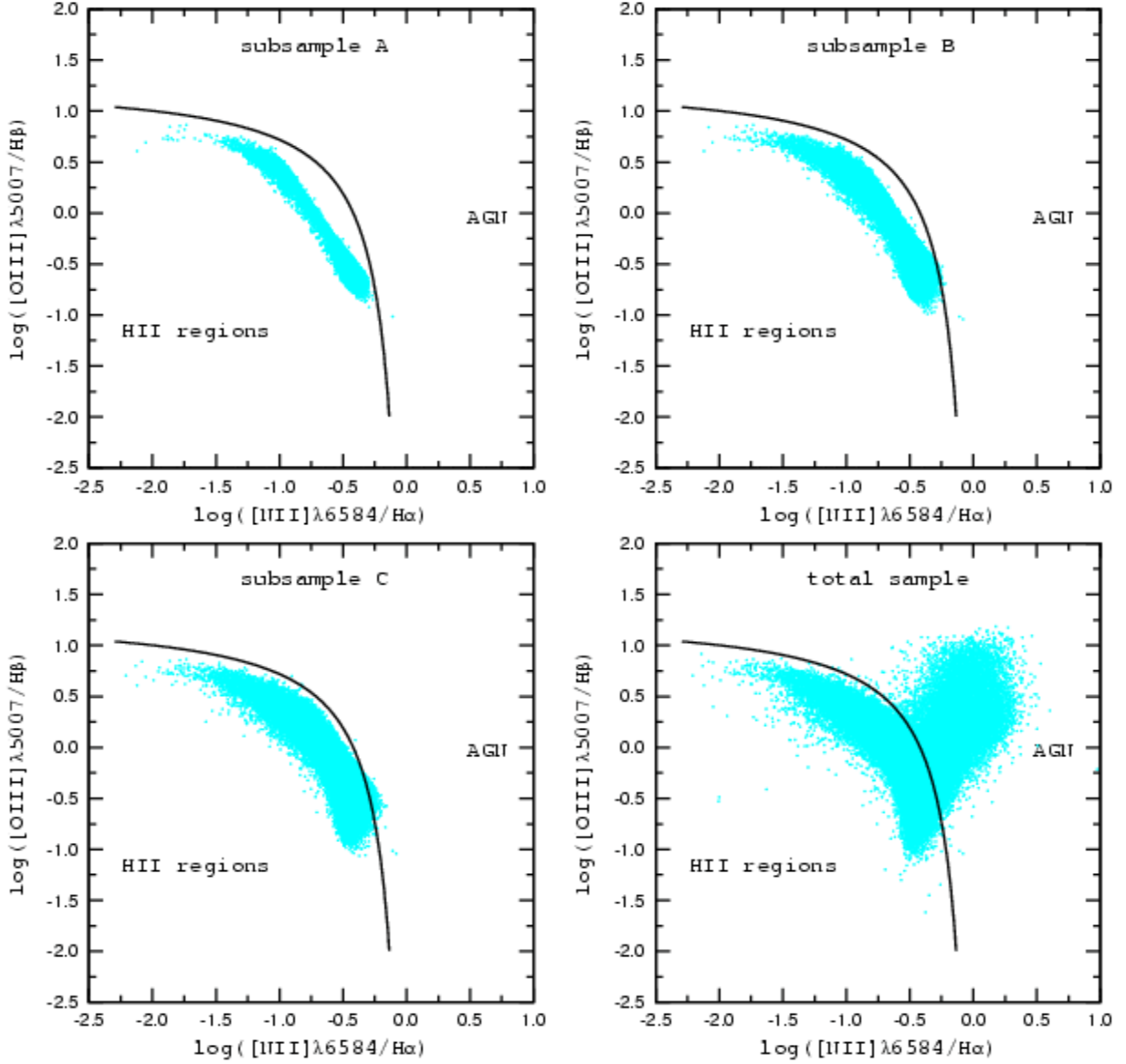}}
\caption{ 
The [OIII]$\lambda$5007)/H$_{\beta}$ vs [NII]$\lambda$6584)/H$_{\alpha}$ 
diagram for subsamples A, B, and C and the total sample.
The solid line shows the dividing line between H\,{\sc ii} regions 
ionized by star clusters and AGNs \citep{kauffmannetal2003}.
}
\label{figure:seagull}
\end{center}
\end{figure}

Does selecting objects in such a way give us only 1) star-forming galaxies 
and 2) with accurate line flux measurements? We can test the reliability 
of our selection criterion concerning the first point by appealing to the 
[OIII]$\lambda$5007/H${\beta}$ vs [NII]$\lambda$6584/H${\alpha}$ 
diagram which is often used to distinguish between  
star-forming galaxies and AGNs.
Fig.\ref{figure:seagull} shows such diagrams for subsamples A, B, C and 
for the total sample. The solid line, taken from \citet{kauffmannetal2003}, 
shows the dividing line 
betweeen H\,{\sc ii} regions ionized by star clusters and AGNs.
Fig.\ref{figure:seagull} clearly shows that, while the 
total sample contains both star-forming galaxies and AGNs (lower right 
panel), our criterion does select out subsamples 
containing only star-forming galaxies.

\begin{figure}
\begin{center}
\resizebox{1.00\hsize}{!}{\includegraphics[angle=0]{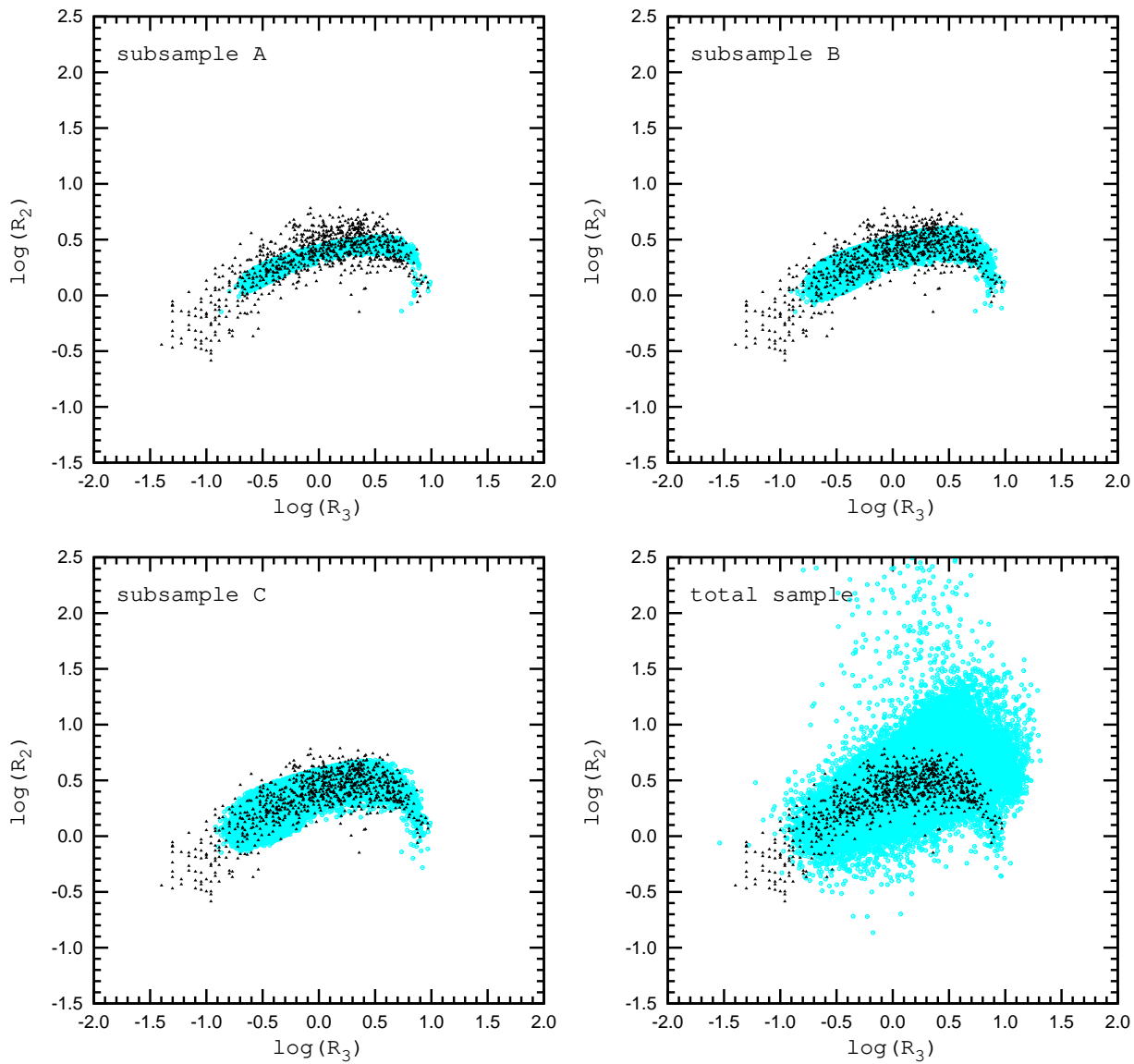}}
\caption{ 
The R$_3$ -- R$_2$  diagram for subsumples A, B, and C and the total sample.
The SDSS objects are shown by gray (light-blue in the electronic version) circles.
The H\,{\sc ii} regions 
in nearby galaxies from the compilation of \citet{pilyuginetal2004} 
are shown by black triangles.  
}
\label{figure:r3r2}
\end{center}
\end{figure}

Concerning the second point, 
Fig.\ref{figure:r3r2} shows the classical R$_3$ -- R$_2$  
diagram for subsamples A, B, and C and the total sample. Each galaxy is 
plotted as a gray (light blue in the electronic version) open circle.
For comparison, H\,{\sc ii} regions 
in nearby galaxies with accurate line flux measurements, 
from the compilation of \citet{pilyuginetal2004},
 are shown as black filled circles. Inspection of 
Fig.\ref{figure:r3r2} shows that the selected subsamples of galaxies 
occupy the same area in the R$_3$ -- R$_2$ diagram as the 
H\,{\sc ii} regions in nearby galaxies with accurate measurements, 
while the total sample covers a considerably larger area. Evidently, our 
selection criterion picks out objects which have line intensities 
that are in agreement with the well-measured line intensities of HII regions 
in nearby galaxies. 

\begin{figure}
\begin{center}
\resizebox{0.75\hsize}{!}{\includegraphics[angle=0]{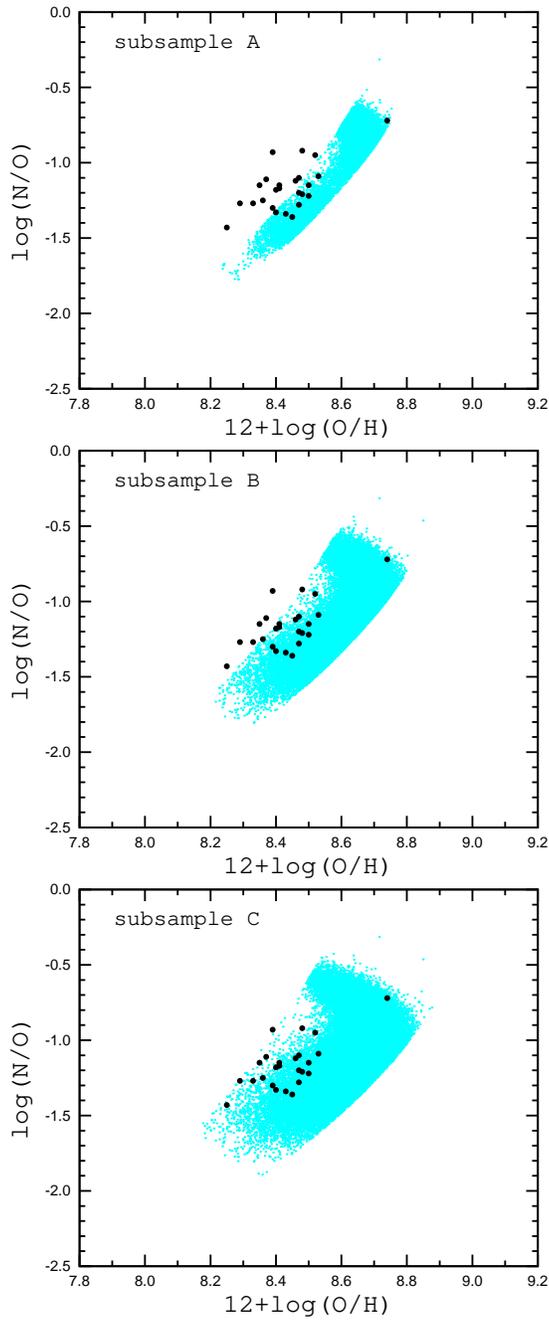}}
\caption{ 
The O/H - N/O  diagram for subsumples A, B, and C.
The SDSS objects are shown by gray (light-blue in the electronic version) open circles.
The H\,{\sc ii} regions in nearby galaxies with measured
electron temperatures 
\citep{bresolin2007,bresolinetal2009b} 
are shown by black filled circles.  
}
\label{figure:ohno}
\end{center}
\end{figure}

We next investigate the O/H -- N/O diagram (Fig.\ref{figure:ohno}).
As before, the SDSS galaxies 
from subsamples A, B, and C are shown as gray (light-blue 
in the electronic version) open circles, and  
the H\,{\sc ii} regions in nearby galaxies with recent precise measurements of 
electron temperatures \citep{bresolin2007,bresolinetal2009a} are shown 
as black filled circles. We see that the selected SDSS galaxies and 
the  H\,{\sc ii} regions in nearby galaxies with precise measurements 
lie in the same general region in the O/H -- N/O diagram. However, the 
selected SDSS galaxies are shifted systematically toward lower N/O ratios. 
This slight systematic 
shift (of $\sim$ 0.2-0.3 dex in log N/O) can be understood 
in the following way. 
The selection criteria pick out galaxies with strong ongoing star formation,
resulting in a higher  
star formation rate in the selected SDSS galaxies as compared 
to normal galaxies with the same mass.
It has long been known that such a star formation burst would 
cause a temporary 
decrease of the N/O ratio in the galaxy 
\citep[e.g.][]{pilyugin1992,pilyugin1993}.
%than average for galaxies of a given mass.
%It is known for a long time \citep[e.g.][]{pilyugin1992,pilyugin1993} that  
%the star formation burst results in the temporary 
%decrease of the N/O ratio in a galaxy. 
%This seems to be a reason of systematic offset between the comparison 
%sample and selected subsamples of the SDSS galaxies.

\begin{figure}
\begin{center}
\resizebox{1.00\hsize}{!}{\includegraphics[angle=0]{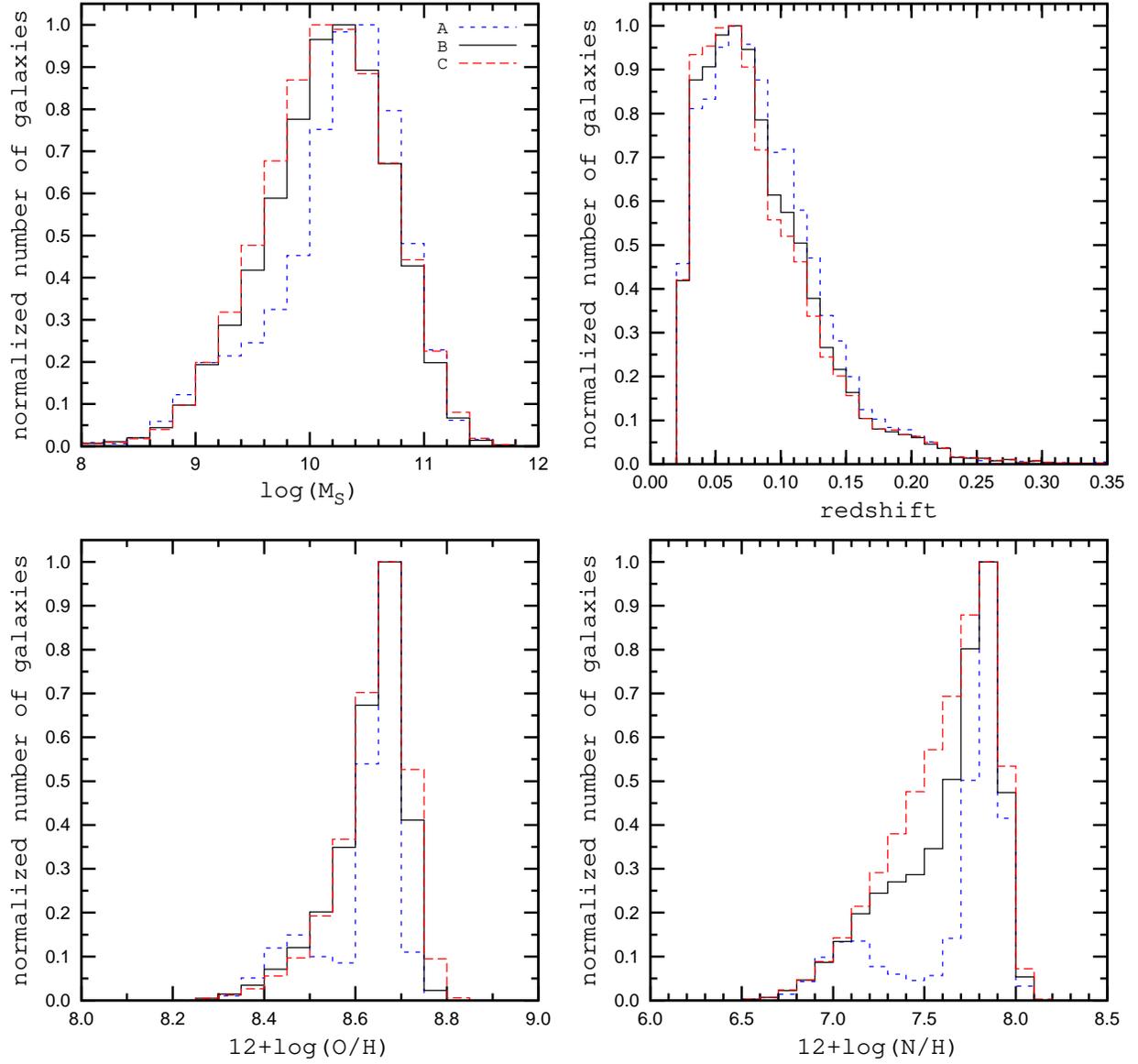}}
\caption{
Normalized histograms of observed and derived properties for SDSS galaxies in 
subsamples A (short-dashed line), B (solid line), 
and C (long-dashed line).
}
\label{figure:gist}
\end{center}
\end{figure}

\begin{figure}
\begin{center}
\resizebox{0.60\hsize}{!}{\includegraphics[angle=0]{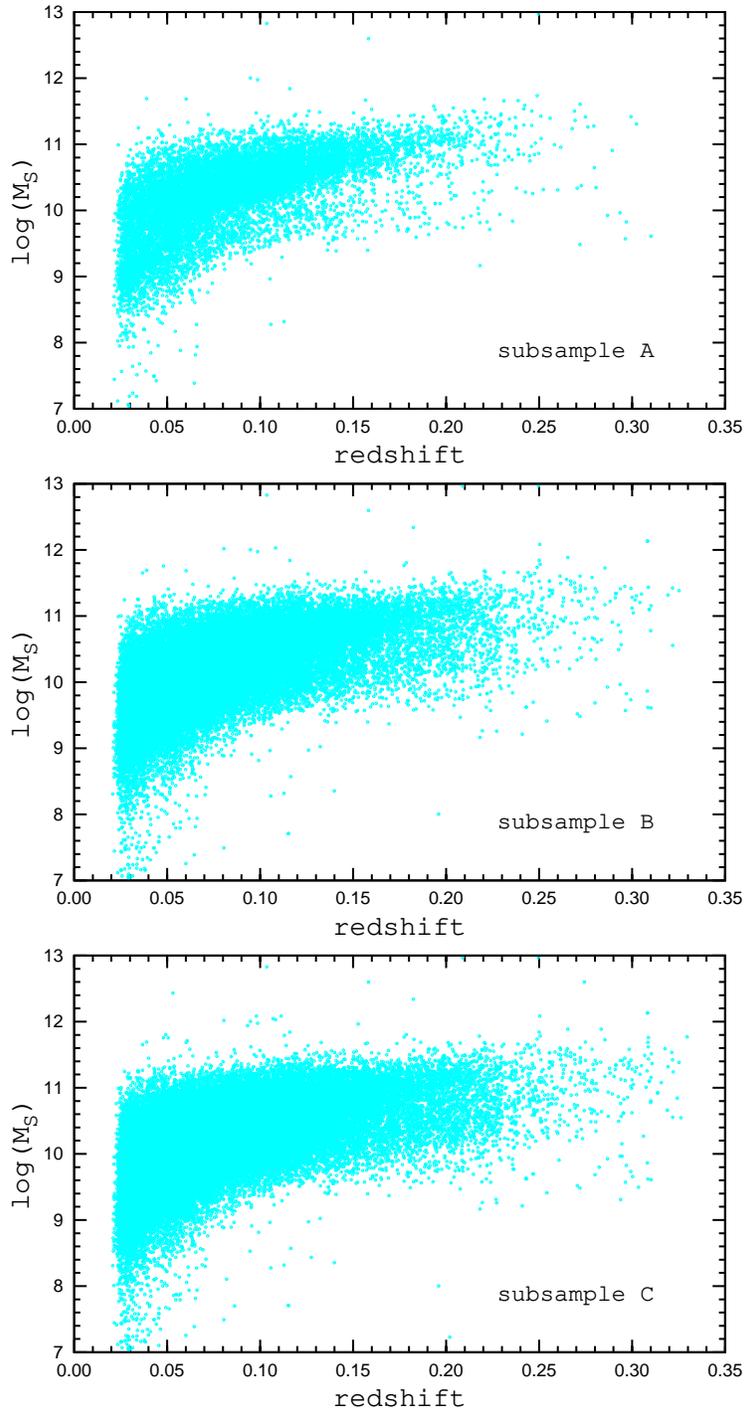}}
\caption{
The redshift -- galaxy stellar mass diagram for  
subsamples A, B, and C.
}
\label{figure:zms}
\end{center}
\end{figure}

We now compare the properties of the galaxies in subsamples A, B, and C.
The histograms of galaxy stellar masses, 
redshifts, oxygen and nitrogen abundances 
for the three subsamples are shown in 
Fig.\ref{figure:gist}.  The locations of the galaxies in 
the three subsamples in the redshift -- galaxy stellar mass diagram 
are presented in Fig.\ref{figure:zms}. 
Examination of the distributions of the parameters in the three subsamples 
reveals that, while they are somewhat similar for subsamples B and C,  
the distribution for subsample A differs significantly. 
In particular, there is a shift toward higher masses (by about 0.2 dex)  
of the galaxy mass distribution 
in subsample A relative to  
the other two subsamples. There is also a shift toward higher 
nitrogen abundances (by about 0.1 dex) of 
galaxies in subsample A relative to subsamples B and C
(see Fig.\ref{figure:gist}). 
These shifts seem to be caused by a too restrictive selection criterion.
The selection condition for subsample A,  
$\Delta$log(N/H)$_{max}$ $\le$ 0.05, appears to be 
too constraining, eliminating too many 
galaxies and, as a consequence, causing a selection effect. 
%which can be understood 
%in the following way.
%Subsample A contains preferentially the most massive galaxies.
%Those will have undergone the most complete chemical evolution. 
%Since there is a time delay between oxygen and nitrogen enrichment, 
%the most massive systems may be those where 
%the most important period of star formation occurred furthest in the past, 
%and where the nitrogen enrichment is closest to being complete, causing a  
%nitrogen enhancement as compared to galaxies in the other two 
%subsamples.   
In order not to bias our results, 
%towards the most massive systems,  
we will consider, in the remainder of the paper, 
subsample B as the basic subsample, while subsample 
C will be used as a control subsample.

\section{THE REDSHIFT EVOLUTION OF OXYGEN AND NITROGEN ABUNDANCES}

We investigate here the changes in oxygen and nitrogen abundances with galaxy 
stellar mass and redshift. 
Examination of Fig.\ref{figure:gist} and Fig.\ref{figure:zms} shows that such 
a study is justified only in the range of stellar galaxy masses from 
$\sim$ 10$^{9.5}$M$_\sun$ to $\sim$ 10$^{11.5}$M$_\sun$, where the number 
of galaxies is large enough to give good statistics.

\subsection{The z--M$_S$--O/H relation}

\begin{figure}
\begin{center}
\resizebox{0.80\hsize}{!}{\includegraphics[angle=0]{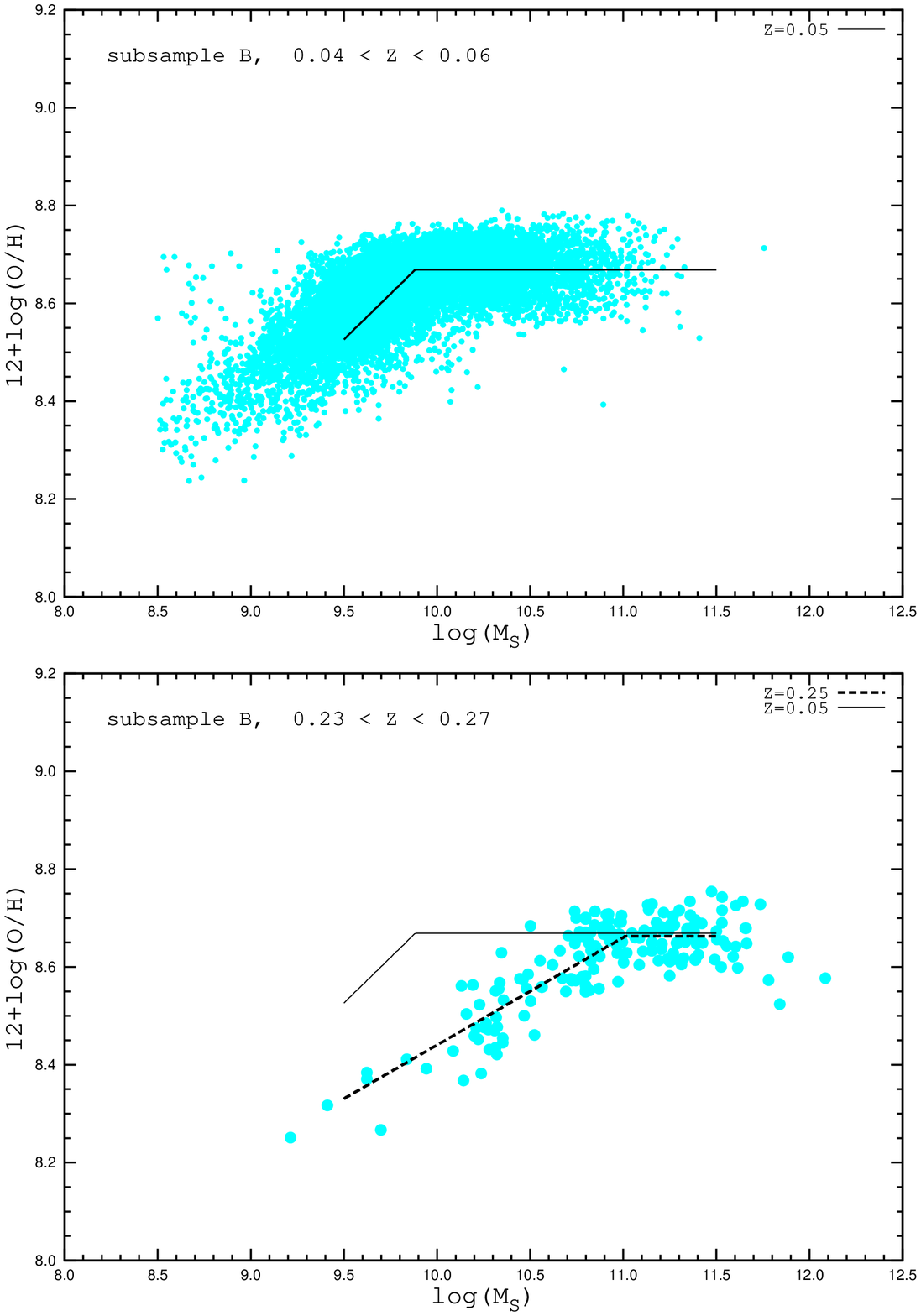}}
\caption{
The oxygen abundance -- galaxy stellar mass diagrams for subsample B 
at $z$=0.05 (upper panel) and $z$=0.25 (lower panel). 
The Z-M$_S$-O/H relation (Eq.\ref{equation:yzmohb}) 
is shown by the solid line  for $z$=0.05 
and by the dashed line for $z$=0.25.
}
\label{figure:zmohb}
\end{center}
\end{figure}

We first discuss the evolution of oxygen abundances with galaxy 
stellar masses. 
The upper panel of Fig.\ref{figure:zmohb} shows the oxygen abundances 
of galaxies in subsample B, with redshifts in the range 0.04 $<$ $z$ $<$ 0.06, 
as a function of galaxy stellar mass. 
The lower panel of Fig.\ref{figure:zmohb} shows the same diagram, but for 
galaxies in a higher redshift range, 0.23 $<$ $z$ $<$ 0.27.
We note that our M$_S$--O/H diagram is very similar to that of  
\citet{erbetal2006} (see their Fig.3). This may not 
be surprising since those authors also used a N$_2$ calibration to  
estimate the oxygen abundances of their SDSS galaxies.
Fig.\ref{figure:zmohb} shows that the oxygen abundance increases 
with increasing galaxy stellar mass up to a value M$_S^*$. 
For galaxies with M$_S$ $>$ M$_S^*$, the oxygen abundance becomes  
constant. Comparison of the upper and lower panels of Fig.\ref{figure:zmohb}
reveals that the value of M$_S^*$ depends on redshift, 
shifting to higher values at higher redshifts. 

To be more quantitative, we approximate 
the change in oxygen abundance with redshift and galaxy stellar mass 
by the following
 ``redshift -- galaxy stellar mass -- oxygen abundance'' z--M$_S$--O/H 
relation:
\begin{eqnarray}
       \begin{array}{llll}
12+\log(O/H) & = & a_1 + a_2\,z + (a_3 + a_4\,z) \log(\frac{M_S}{M_S^*}), &  M_S \leq M_S^* \\
12+\log(O/H) & = & a_1 + a_2\,z ,                                         &  M_S \geq M_S^* \\
\log(M_S^*/M_{\sun})   & = & a_5 + a_6\ z                                           &                  
       \end{array}
\label{equation:ygeneral}
\end{eqnarray}
Fitting the data for galaxies in subsample B gives the following:
\begin{eqnarray}
       \begin{array}{llll}
12+\log(O/H) & = & 8.67 - 0.027\,z + (0.41 - 0.76\,z) \log(\frac{M_S}{M_S^*}), &  M_S \leq M_S^* \\
12+\log(O/H) & = & 8.67 - 0.027\,z ,                                           &  M_S \geq M_S^* \\
\log(M_S^*/M_{\sun})  & = & 9.60 + 5.65\,z                                              &
       \end{array}
\label{equation:yzmohb}  
\end{eqnarray}
The derived z--M$_S$--O/H relation for subsample B 
is shown in Fig.\ref{figure:zmohb} by the solid 
line for $z$=0.05, and by the dashed line for $z$=0.25.

\begin{figure}
\begin{center}
\resizebox{0.80\hsize}{!}{\includegraphics[angle=0]{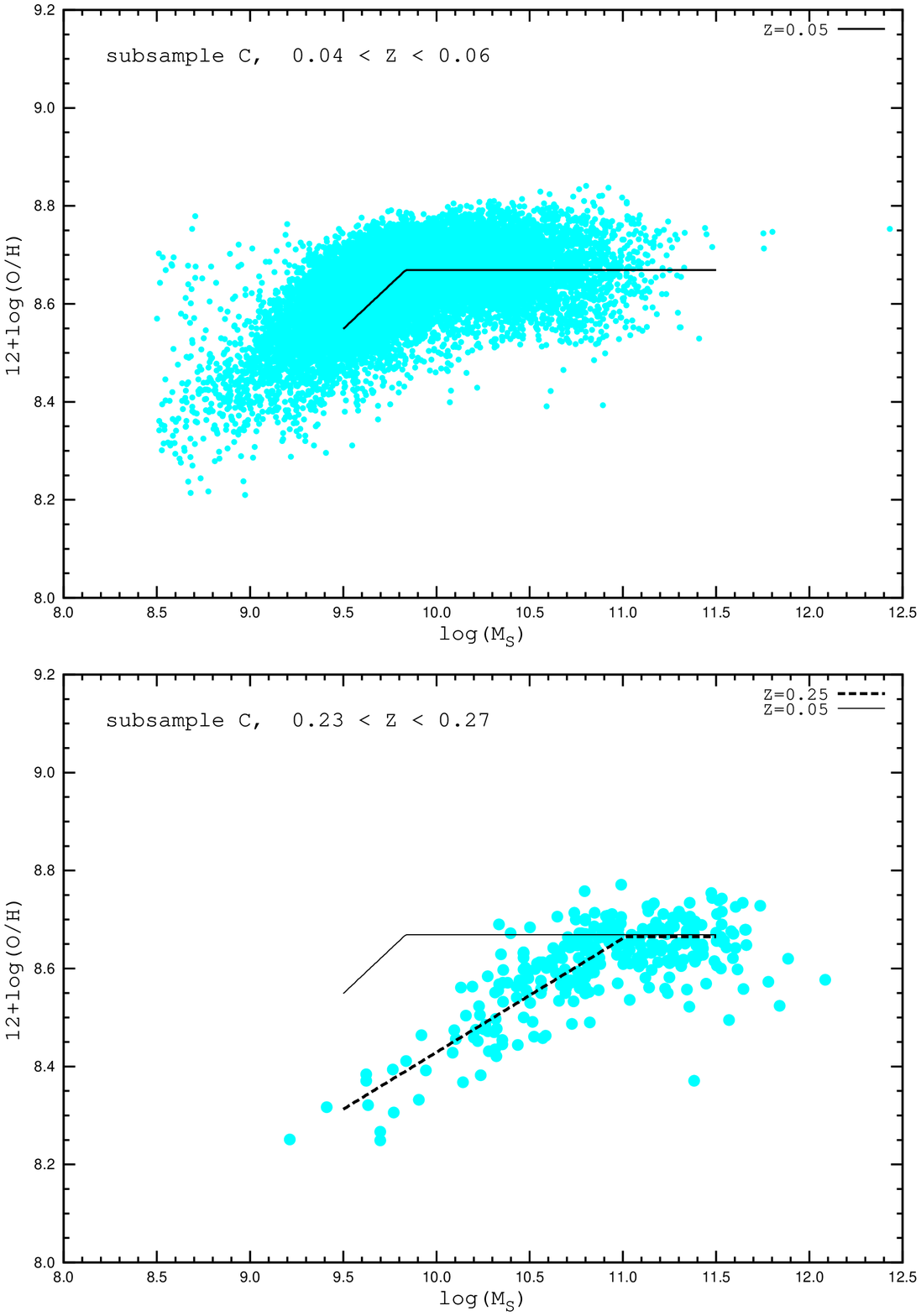}}
\caption{
The oxygen abundance -- galaxy stellar mass diagrams for subsample C 
at $z$=0.05 (upper panel) and $z$=0.25 (lower panel). 
The Z-M$_S$-O/H relation (Eq.\ref{equation:yzmohc}) 
is shown by the solid line  for $z$=0.05 
and by the dashed line for $z$=0.25.
}
\label{figure:zmohc}
\end{center}
\end{figure}

To test the robustness of 
the derived z--M$_S$--O/H relation, we have also analyzed 
subsample C in a similar way.
The data for galaxies in subsample C is shown in Fig.\ref{figure:zmohc}.
Fitting the subsample C data gives the following: 
\begin{eqnarray}
       \begin{array}{llll}
12+\log(O/H) & = & 8.67 - 0.020\,z + (0.39 - 0.63\,z) \log(\frac{M_S}{M_S^*}), &  M_S \leq M_S^* \\
12+\log(O/H) & = & 8.67 - 0.020\,z ,                                           &  M_S \geq M_S^* \\
\log(M_S^*/M_{\sun})  & = &  9.54 + 5.90\,z                                             &
       \end{array}
\label{equation:yzmohc}  
\end{eqnarray}
The z--M$_S$--O/H relation for subsample C 
is shown in Fig.\ref{figure:zmohc} by the solid 
line for $z$=0.05, and by the dashed line for $z$=0.25.
Comparison between Fig.\ref{figure:zmohb} and Fig.\ref{figure:zmohc} 
shows that the obtained z--M$_S$--O/H relations for 
subsamples B and C are very similar, so that our results appear to be robust.
The derived relations should approximate 
well the redshift evolution 
of oxygen abundances for $z$ $\leq$ 0.25,  
corresponding to lookback times of up to $\sim$ 3 Gyr. 
Beyond $z$ = 0.25, 
the number of galaxies in each subsample becomes very small 
(see Fig.\ref{figure:zms}) and the redshift evolution of 
abundances cannot be studied with enough statistics.
To discuss abundance evolution, we will therefore restrict 
ourselves to objects with  $z$ $\leq$ 0.25.  
At the low-redshift end, 
to reduce errors from aperture effects and following 
the recommendations of \citet{kewleyetal2005}, 
we will limit ourselves to objects with 
redshifts $z$ $>$ 0.04. Objects 
with 0.04 $<$ $z$ $<$ 0.06 will be considered as representative of 
the present-day epoch.

Fig.\ref{figure:egist} shows the distribution of the  
deviations of the oxygen abundances 
from the derived z--M$_S$--O/H  relation for both subsamples B and C.
It is seen 
that the scatter of oxygen abundances about the 
z--M$_S$--O/H  relation is slightly 
larger for galaxies in subsample C than in subsample B. 
This effect can also be seen by comparing Fig.\ref{figure:zmohb} to  
Fig.\ref{figure:zmohc}.  
Nevertheless, it can be said that, for both subsamples of 
galaxies, the oxygen abundances follow reasonably well 
the derived  z--M$_S$--O/H  relation. The 
mean deviation is small, being $\sim$ 0.06 dex.
We thus conclude that 
the derived relations can be used, in principle, to obtain 
a rough estimate of the metallicities of SDSS galaxies.

\subsection{The metallicity plateau and the oxygen enrichment as a function of galaxy stellar mass}

One of the most remarkable features of the   
M$_S$--O/H relation at the current epoch  
is the metallicity 
plateau at high galaxy stellar masses. 
For the sake of definiteness, 
we will discuss the results for subsample B. 
The oxygen abundance increases 
with galaxy stellar mass until M$_S$ = 10$^{9.9}$M$_\sun$, but then remains 
approximately constant, equal to
12+log(O/H)=8.67, 
in galaxies with M$_S$ $\ga$ 10$^{9.9}$M$_\sun$ 
(see the upper panel of Fig.\ref{figure:zmohb}). 
What is the physical meaning of such a plateau? 
The observed oxygen abundance in a galaxy is defined by its astration level 
or gas mass fraction $\mu$, and by the mass exchange between a galaxy and its 
environment \citep{searlesargent1972,pagel1997}.
It is believed that galactic winds do not play a significant role in 
the chemical evolution of large spiral galaxies 
\citep{garnett2002,tremontietal2004,dalcanton2007}
and that the rate of gas infall onto 
the disk decreases exponentially with time
\citep{matteuccifrancois1989,pilyuginedmunds1996a,pilyuginedmunds1996b,caluraetal2009}. 
The present-day location of a system 
in the $\mu$--O/H diagram is then 
governed by its evolution in the recent past, and 
is only weakly dependent of its evolution on long time-scales
\citep{pilyuginferrini1998}. Therefore  
the observed oxygen abundance in a large spiral galaxy is mainly defined by 
its gas mass fraction $\mu$. 
The presence of a metallicity plateau in the M$_S$--O/H diagram 
at all redshifts implies then   
that the high-mass SDSS galaxies have very similar gas mass fractions.  

The luminosity--central metallicity relation for nearby spiral galaxies 
also shows a plateau at high luminosities \citep{pilyuginetal2007}. 
This plateau was interpreted as evidence that the gas in the centers of the 
most metal-rich galaxies has been almost completely converted into stars and 
that the oxygen abundance in the centers of the 
most luminous metal-rich galaxies has reached its 
maximum attainable value of 12+log(O/H) $\sim$ 8.87. 
The plateau for oxygen abundances in the SDSS galaxies is at the level 
of 12+log(O/H)=8.67, i.e. 0.2 dex smaller.
The simple model of chemical evolution of galaxies 
predicts that a decrease of $\mu$  by 0.1 results in an 
increase of oxygen abundance by $\sim$0.13 dex, in the range of $\mu$ from 
$\sim$0.50 to $\sim$ 0.05 \citep{pilyuginetal2007}. 
Then, the difference between the maximum attainable value 
of the oxygen abundance and 
the mean oxygen abundance in high-mass SDSS galaxies,  
$\Delta$(log(O/H)) $\sim$ 0.2, corresponds to 
a difference in gas mass fraction $\Delta$$\mu$ $\sim$ 0.15. 
Since the maximum attainable oxygen value corresponds to complete astration, i.e. 
$\mu$ = 0, the mean $\mu$ in high-mass SDSS galaxies at the present 
epoch should be $\sim$ 15\%, with a probable range of $\mu$ 
from $\sim$ 5\% to $\sim$ 25\%. 

Why do our SDSS subsamples not contain galaxies with abundances 
as high as the maximum attainable value? The reason has to do with 
the gas content of the SDSS galaxies in our subsamples. 
As said before, the maximum attainable value of the oxygen abundance corresponds to 
the limiting case where the gas has been completely converted into stars. 
The galaxies in our SDSS subsamples are all characterized by strong emission lines in 
their spectra, meaning that they are undergoing strong starbursts. This requires in turn 
that they contain an appreciable amount of gas. In other words, because our galaxies 
are gas-rich, they have not reached the maximum attainable value of the oxygen 
abundance which requires complete gas exhaustion. If galaxies 
were not gas-rich, they would have weak emission lines, unlikely to be 
measured accurately. 
There exists thus a lower limit on the gas mass fraction for a galaxy 
to have accurately measured lines and to be included in our subsamples.
In that sense, the observed plateau is likely affected by the 
selection criteria. We note however that, even in the most evolved nearby 
spiral galaxies, the H\,{\sc ii} region oxygen abundances 
are generally lower than the maximum attainable value. 
There is furthermore an aperture effect that slightly lowers the metallicities 
observed for SDSS galaxies. 
The maximum attainable value oxygen abundance observed in the 
central part of some 
nearby galaxies is usually derived by linearly fitting the variations of 
H {\sc ii} region abundances with galactocentric distance, and 
extrapolating to R=0. In the distant SDSS galaxies, because one  
fiber includes many H\,{\sc ii} regions that show decreasing 
metallicities towards larger galactocentric distances, 
the oxygen abundances are slightly diluted when averaged over a large region. 

In the lower panel of  Fig.\ref{figure:zmohb}, we compare the M$_S$--O/H relation 
for local  ($z$ $\approx$ 0.05) galaxies (solid line) with that for distant  
($z$ $\approx$ 0.25) ones (dashed line). 
Three features are to be noted. First, it can be seen that, for the galaxies of 
highest stellar masses, those with masses $\ga$ 10$^{11}$M$_\sun$, 
the plateau value of the oxygen abundance does not change in 
the redshift interval from 0.05 to 0.25. 
 This implies that the galaxies of highest mass in our SDSS subsamples have reached their highest 
astration level some 3 Gyr ago, and have been somewhat "lazy" in their 
evolution afterwards. 
Second, for lower mass galaxies with stellar masses $\la$ 10$^{11}$M$_\sun$,  
it is seen that the value of the oxygen enrichment during the last 3 Gyr increases 
with decreasing galaxy mass, in the mass interval from 10$^{11}$M$_\sun$ to  
10$^{9.9}$M$_\sun$, as shown by the widening gap between the solid 
and dashed curves toward lower masses.  
At M$_S$=10$^{9.9}$M$_\sun$, there is a difference $\Delta$log(O/H) $\sim$ 0.25 
between a local galaxy and one at redshift 0.25. For galaxies with 
M$_S$ $\leq$ 10$^{9.9}$M$_\sun$, the oxygen enrichment during the last 3 Gyr 
slightly decreases with decreasing galaxy mass, the slope of the solid line 
being slightly steeper than that of the dashed line. Third, the value of 
M$_S^*$, the mass where the oxygen 
abundance becomes constant with galaxy stellar mass,
 is redshift-dependent, becoming higher at larger redshifts. 

In summary, analysis of the z--M$_S$--O/H  relation has led to the following
main conclusions. \\
-- The galaxies of highest masses, those with M$_S$ $\ga$ 10$^{11}$M$_\sun$,
 have reached their highest astration level in 
the past and have not had an appreciable oxygen abundance enrichment  
during the last $\sim$ 3 Gyr.  \\
-- The mean value of the oxygen enrichment during the last 3 Gyr in galaxies with 
stellar masses in the range from 10$^{10}$M$_\sun$ to 10$^{11}$M$_\sun$  
is $\Delta$(log(O/H)) $\sim$ 0.11, with 
$\Delta$(log(O/H)) $\sim$ 0.23 at 10$^{10}$M$_\sun$ and 
$\Delta$(log(O/H)) =0 at 10$^{11}$M$_\sun$.  

The above picture will now be put to the test through  the study of the redshift evolution of  
nitrogen abundances, which we consider next. 

\subsection{The z--M$_S$--N/H relation}

We proceed in the same way as in our analysis of the oxygen abundances. 
The upper panel of Fig.\ref{figure:zmnhb} shows the nitrogen abundances 
of the local galaxies in subsample B, with redshifts 0.04 $<$ $z$ $<$ 0.06,  as a function 
of galaxy stellar mass. 
The lower panel of Fig.\ref{figure:zmnhb} shows the same diagram, but for 
more distant galaxies, with redshifts 0.23 $<$ $z$ $<$ 0.27.
Fig.\ref{figure:zmnhb} shows that the general behavior of nitrogen abundances 
with redshift and galaxy stellar mass is similar to that of oxygen abundances.
The nitrogen abundance increases 
with increasing galaxy stellar mass up to a value M$_S^*$ of the stellar 
mass. Then, for galaxies with M$_S$ $>$ M$_S^*$, the nitrogen abundance remains 
approximatively constant, reaching a plateau. The value of M$_S^*$ is redshift-dependent, 
becoming higher at larger redshifts. 

The change of nitrogen abundance with redshift and galaxy stellar mass 
can be approximated by a  z--M$_S$--N/H relation, similar to the one for oxygen 
abundances. 
Fitting the data for galaxies in subsample B results in the following 
relation:
\begin{eqnarray}
       \begin{array}{llll}
12+\log(N/H) & = & 7.89 - 0.110\,z + (0.99 - 1.40\,z) \log(\frac{M_S}{M_S^*}), &  M_S \leq M_S^* \\
12+\log(N/H) & = & 7.89 - 0.110\,z ,                                           &  M_S \geq M_S^* \\
\log(M_S^*/M_{\sun})  & = & 9.97 + 4.92\,z                                              &
       \end{array}
\label{equation:yzmnhb}  
\end{eqnarray}
The derived relation is shown in Fig.\ref{figure:zmnhb} by a solid 
line for $z$=0.05, and by a dashed line for $z$=0.25.

Again, to test the robustness of the obtained z--M$_S$--N/H relation, 
we have examined subsample C in a similar way.
The data for galaxies in subsample C are plotted in Fig.\ref{figure:zmnhc}.
Fitting those data gives:
\begin{eqnarray}
       \begin{array}{llll}
12+\log(N/H) & = & 7.88 - 0.027\,z + (0.85 - 1.14\,z) \log(\frac{M_S}{M_S^*}), &  M_S < M_S^* \\
12+\log(N/H) & = & 7.88 - 0.027\,z ,                                           &  M_S > M_S^* \\
\log(M_S^*/M_{\sun})  & = & 10.04 + 5.03\,z                                             &
       \end{array}
\label{equation:yzmnhc}  
\end{eqnarray}
The derived z--M$_S$--N/H relation is shown in Fig.\ref{figure:zmnhc} by a solid 
line for $z$=0.05, and by a dashed line for $z$=0.25.
Comparison of Figures \ref{figure:zmnhb} and \ref{figure:zmnhc} 
shows that they are very similar and that the derived z--M$_S$--N/H relation is robust.

While the general behavior of nitrogen abundances with redshift and 
galaxy stellar mass is quite similar to that of oxygen abundances 
(compare Figs.\ref{figure:zmohb} and \ref{figure:zmnhb}), there are significant 
differences, due to the different production mechanisms of these two elements. 
The dependence of the nitrogen abundance on galaxy stellar  
mass for M$_S$ $<$ M$_S^*$ is considerably steeper than 
that for oxygen abundances. 
This is caused by the fact that at oxygen abundances higher than about 
12+log(O/H)=8.3, the metallicity-dependent nitrogen production by 
intermediate-mass stars starts to dominate.  
Another remarkable difference concerns M$_S^*$, the mass which marks 
the transition from the linear regime to the plateau regime: at all redshifts, 
 M$_S^*$ is shifted towards higher values in the M$_S$--N/H diagram as compared to in the 
M$_S$--O/H diagram. Thus,  logM$_S^*$ is equal to 10.2 and 11.2 at z = 0.05 
and z = 0.25 respectively in the M$_S$--N/H diagram, as compared to 9.9 
and 11.0 in the M$_S$--O/H diagram.     

We now attempt to understand this M$_S^*$ shift. The time delay in nitrogen 
production relative to oxygen production plays an important role. 
Examination of Figs.\ref{figure:zmohb} and \ref{figure:zmnhb} shows that 
galaxies with masses  $\ga$ 10$^{11.2}$M$_\sun$ are in the 
plateau regime in both the M$_S$--O/H and in M$_S$--N/H diagrams, at 
both $z$=0.05 and $z$=0.25. This means that those galaxies have not 
undergone appreciable enrichment in both oxygen and nitrogen during the last $\sim$ 3 Gyr.
Evidently, there has not been appreciable star formation 
in those galaxies over the redshift range from $z$=0.25 to $z$=0.05.
 Significant star formation in those galaxies 
has occurred so long ago that stars have returned their 
nucleosynthesis products to the interstellar medium before the epoch corresponding to $z$=0.25.
  
Galaxies with masses between $\sim$ 10$^{11.0}$M$_\sun$  
and $\sim$ 10$^{11.2}$M$_\sun$ do not show an appreciable enrichment  
in oxygen abundance from $z$=0.25 to $z$=0.05 but do show some enrichment 
in nitrogen over this period. This suggests that 
there has not been appreciable star formation 
in those galaxies over the period from $z$=0.25 to $z$=0.05. However, they  
do contain 
stars that were formed before $z$=0.25, but later in comparison to 
the galaxies of highest masses. The massive oxygen-producing stars die 
after a few million years, releasing oxygen in the interstellar medium. 
By contrast, the nitrogen-producing 
intermediate-mass stars have longer lifetimes, and 
so they have not returned nitrogen to the 
interstellar medium before $z$=0.25 because they have not had enough 
time to evolve. 
This also suggests that stars that make a contribution to the nitrogen production 
have lifetimes of a few Gyr.

The galaxies with masses $\la$ 10$^{11.0}$M$_\sun$ show 
enrichment in both oxygen and nitrogen abundances 
after the period corresponding to $z$=0.25. 
This means that appreciable star formation has taken place 
in those galaxies during the last $\sim$ 3 Gyr.
The nitrogen production increases with decreasing galaxy mass from 10$^{11.2}$M$_\sun$ 
to $\sim$ 10$^{10.2}$M$_\sun$ where it 
reaches a value $\Delta$log(N/H) $\sim$ 0.65, then 
it decreases with further decrease of galaxy mass. 

\begin{figure}
\begin{center}
\resizebox{0.80\hsize}{!}{\includegraphics[angle=0]{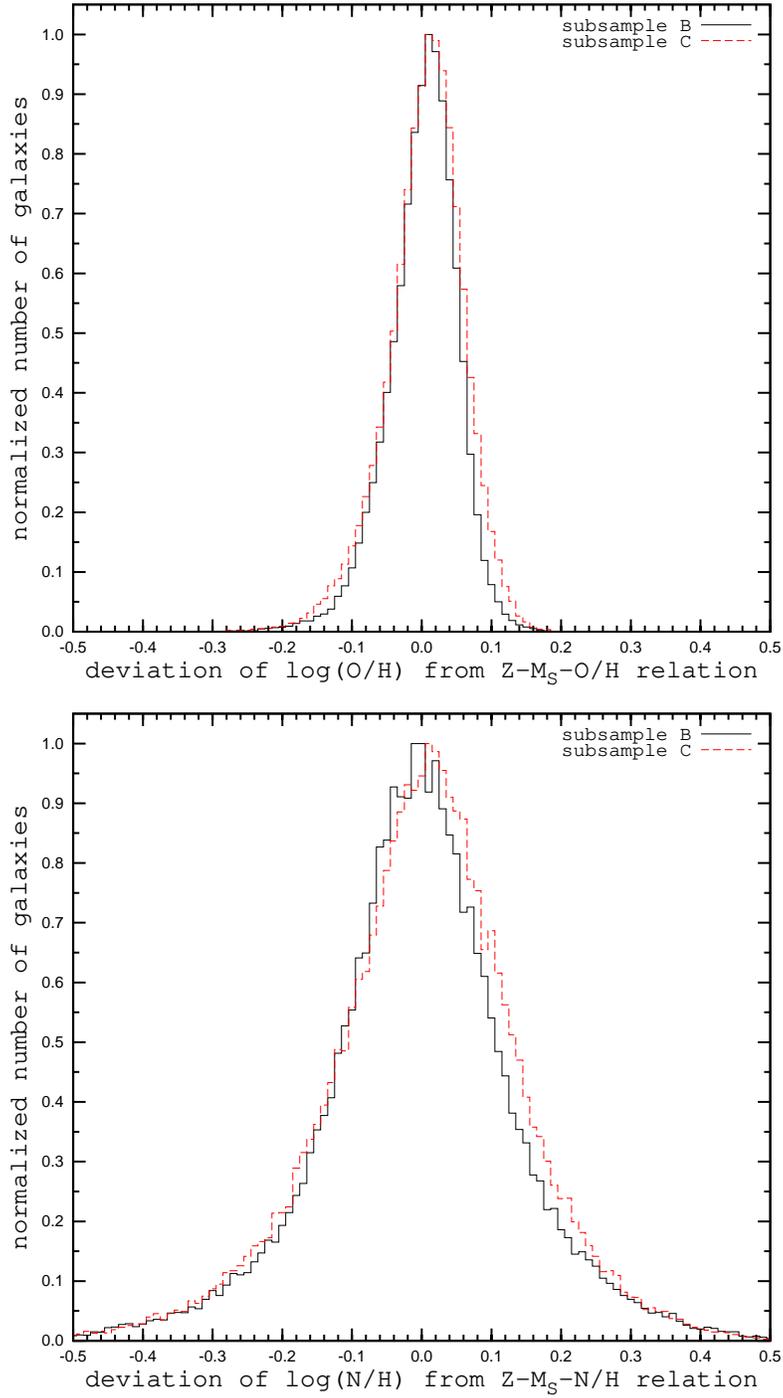}}
\caption{
Normalized histograms of oxygen abundance deviations from the 
Z--M$_S$--O/H relation (upper panel) and nitrogen abundances 
deviations from the  Z--M$_S$--N/H relation (lower panel). 
In both panels, subsample B is shown by the solid line and subsample C by the dashed line.
}
\label{figure:egist}
\end{center}
\end{figure}

\begin{figure}
\begin{center}
\resizebox{0.80\hsize}{!}{\includegraphics[angle=0]{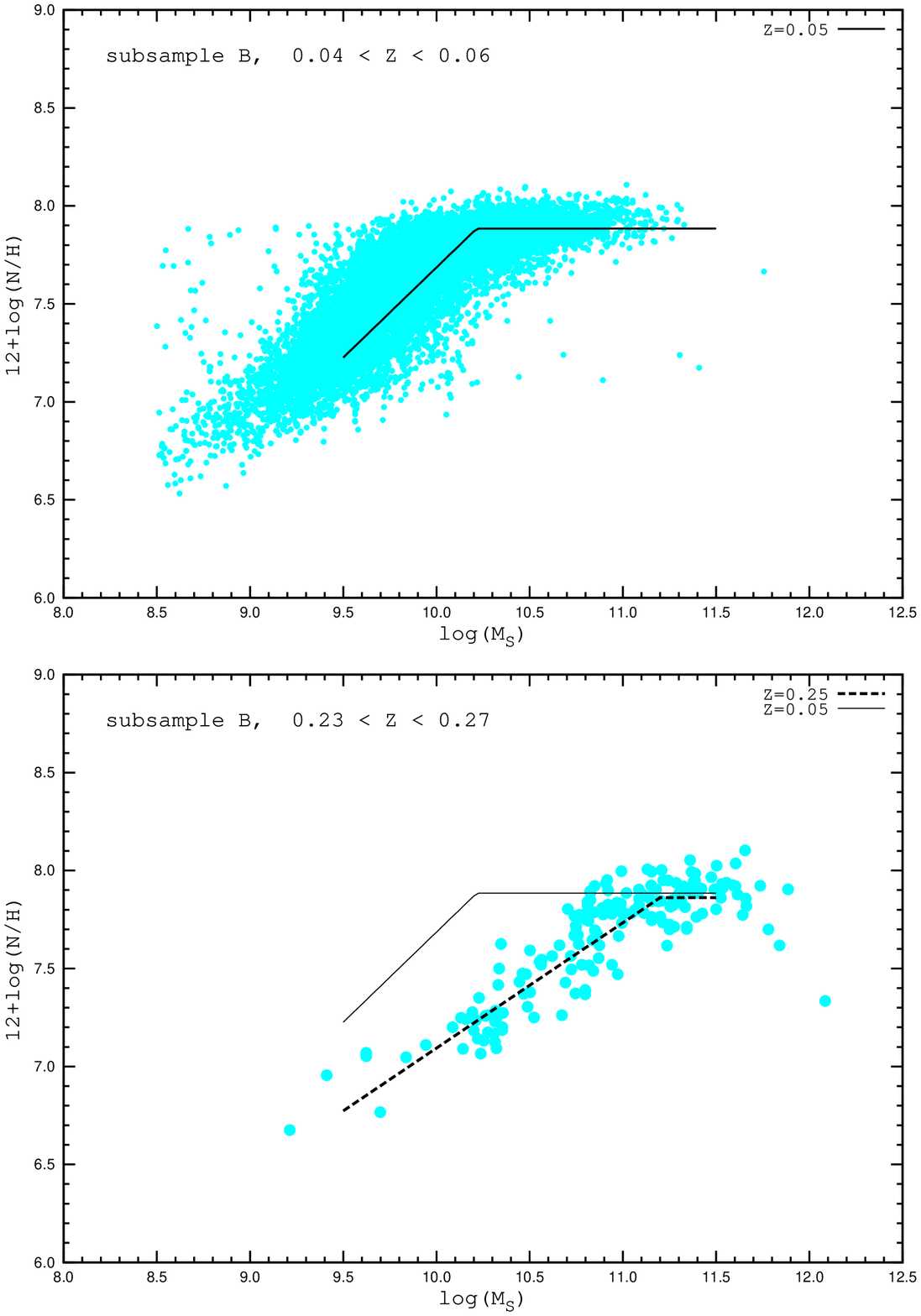}}
\caption{
The nitrogen abundance -- galaxy stellar mass diagrams for subsample B 
at $z$=0.05 (upper panel) and $z$=0.25 (lower panel). 
The Z-M$_S$-N/H relation (Eq.\ref{equation:yzmnhb}) 
is shown by the solid line  for $z$=0.05 
and by the dashed line for $z$=0.25.
}
\label{figure:zmnhb}
\end{center}
\end{figure}

\begin{figure}
\begin{center}
\resizebox{0.80\hsize}{!}{\includegraphics[angle=0]{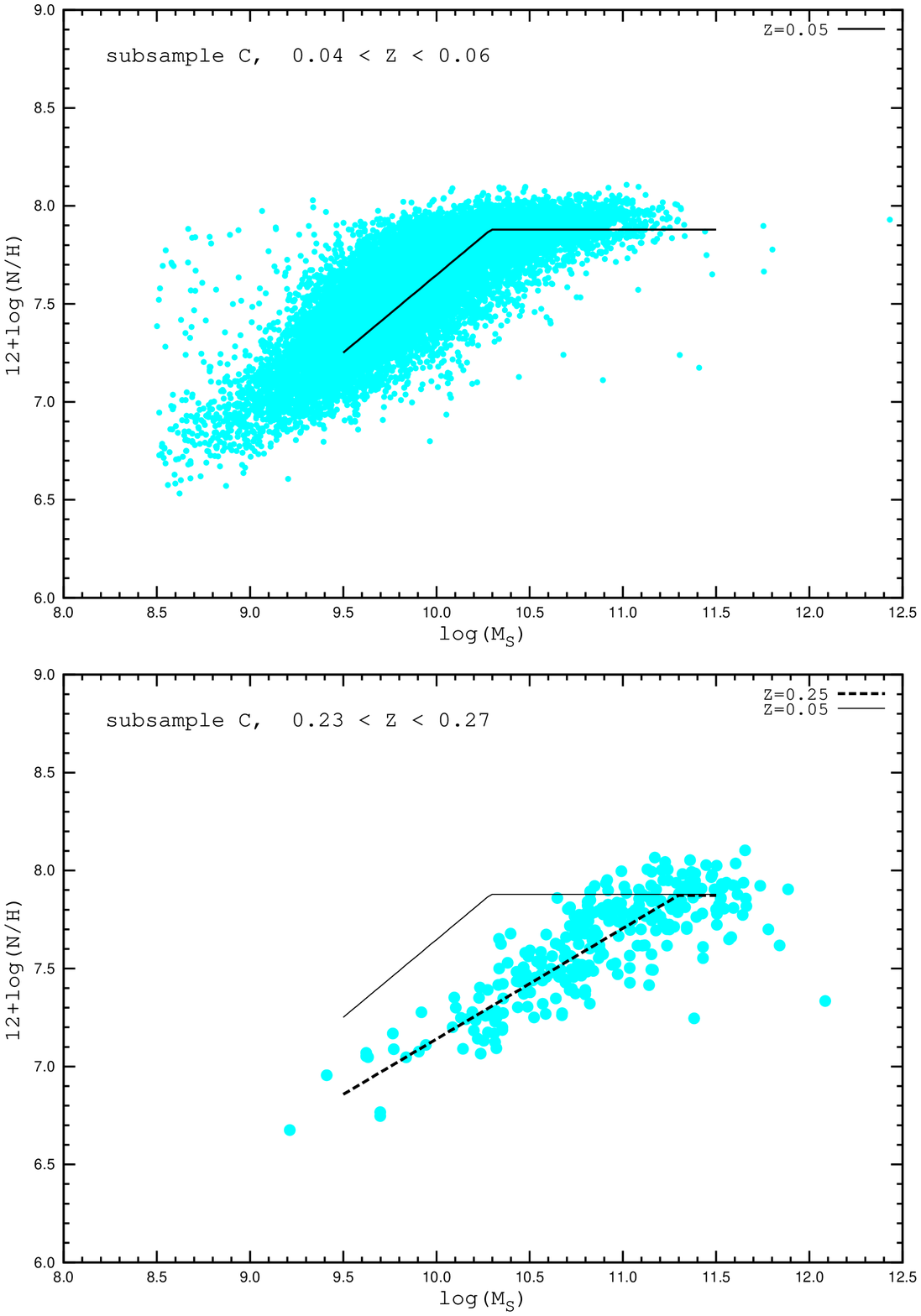}}
\caption{
The nitrogen abundance -- galaxy stellar mass diagrams for subsample C 
at $z$=0.05 (upper panel) and $z$=0.25 (lower panel). 
The Z-M$_S$-N/H relation (Eq.\ref{equation:yzmnhc}), 
is shown by the solid line  for $z$=0.05 
and by the dashed line for $z$=0.25.
}
\label{figure:zmnhc}
\end{center}
\end{figure}

The lower panel of Fig.\ref{figure:egist} shows the deviations of nitrogen 
abundances from the derived z--M$_S$--N/H relation for galaxies from both 
subsamples B (solid line) and C (dashed line). 
Examination of the upper and lower panels of Fig.\ref{figure:egist}  
shows that the lower histogram is broader than the upper one, i.e. that the 
nitrogen abundances show a larger scatter around the  z--M$_S$--N/H relation 
as compared to oxygen abundances around the z--M$_S$--O/H relation.
The mean deviation for nitrogen abundances is  $\Delta$(log(N/H)) $\sim$ 0.15 against   
$\Delta$(log(OH)) $\sim$ 0.06 for oxygen abundances. 
However, nitrogen also spans a significantly larger abundance range,  
from 12+log(N/H) $\sim$ 6.5 to 12+log(N/H) $\sim$ 8.0, 
as compared to oxygen which goes 
only from 12+log(O/H) $\sim$ 8.25 to 12+log(O/H) $\sim$ 8.75. 
Taking into account the differences in range, then the relative 
scatters around the z--M$_S$--N/H and z--M$_S$--O/H relations are comparable. 
It should be emphasized that the larger scatter of nitrogen abundances 
around the z--M$_S$--N/H relation in comparison to the scatter of oxygen abundances 
around the z--M$_S$--O/H relation cannot be attributed to larger errors in 
nitrogen abundance determinations. Indeed, the O/H values are derived from the N/H and N/O 
values, and any error in the nitrogen abundance determination is 
propagated into the 
error in the oxygen abundance determination. Then, if the large scatter of nitrogen abundances 
around the z--M$_S$--N/H  is caused by large errors in the 
nitrogen abundance determinations, then the oxygen abundances would 
show a similar or larger scatter around the z--M$_S$--O/H relation. 
Just the opposite is observed.

\begin{figure}
\begin{center}
\resizebox{1.00\hsize}{!}{\includegraphics[angle=0]{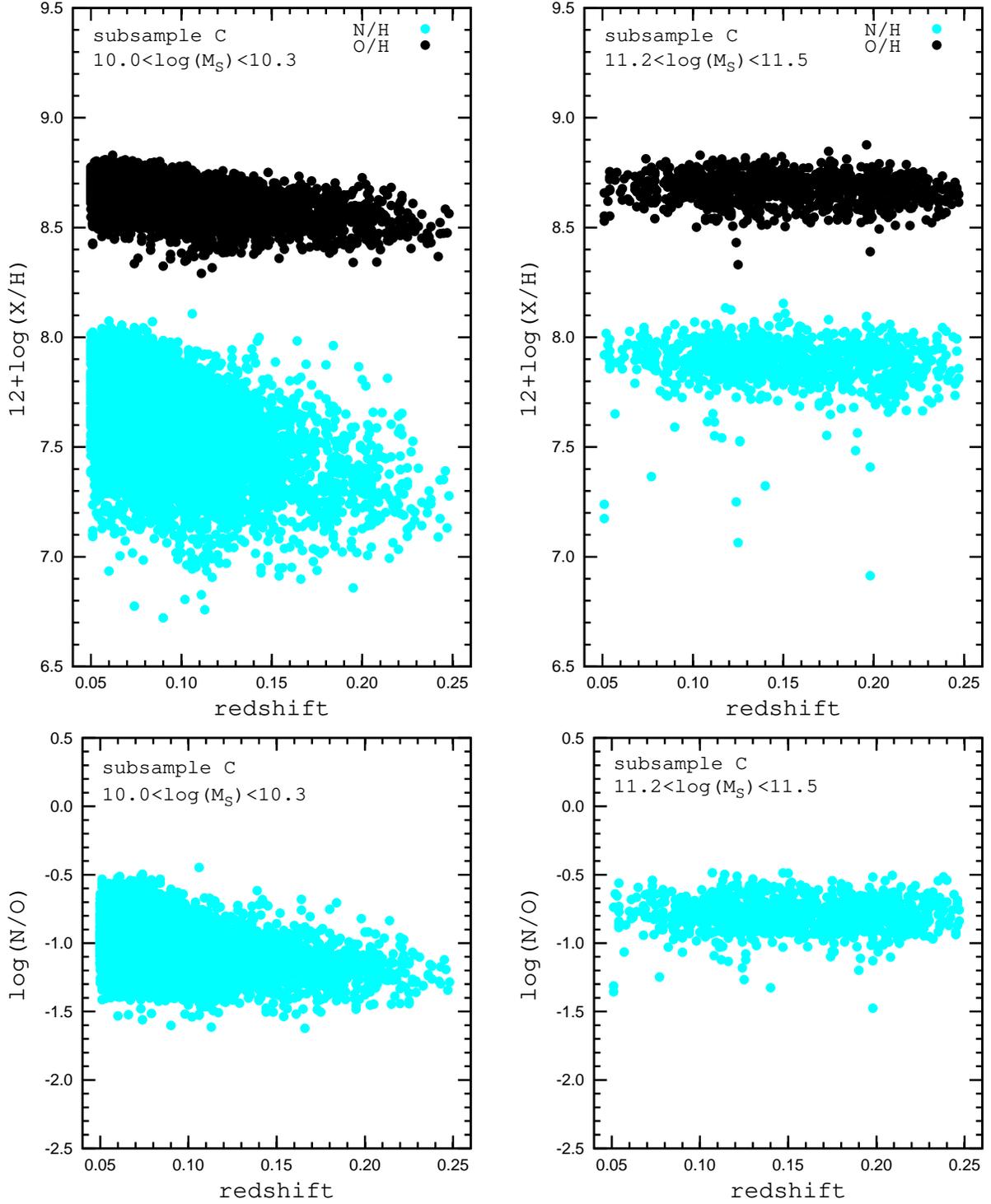}}
\caption{
The oxygen and nitrogen abundances as a function of
redshift for galaxies with masses  10$^{10.0}$M$_\sun$--10$^{10.3}$M$_\sun$ (upper left panel)  
and 10$^{11.2}$M$_\sun$--10$^{11.5}$M$_\sun$ (upper right panel). 
Oxygen abundances are shown by dark (black in the electronic version) 
filled circles, and nitrogen abundances by gray (light-blue in the electronic version) filled circles.
The nitrogen-to-oxygen abundance ratios as a function of redshift for those 
galaxies are shown in the corresponding lower panels.  
}
\label{figure:znod}
\end{center}
\end{figure}

The larger scatter of nitrogen abundances as compared to 
that of oxygen abundances can be understood in the following way. 
The oxygen abundance in a galaxy is mainly defined by its astration level. 
In contrast, the nitrogen abundance is defined not only by the astration level but 
depends also on the star formation history of the galaxy. 
As noted above,  
 a star formation burst results in a temporary 
decrease of the N/O ratio in a galaxy \citep[e.g.][]{pilyugin1992,pilyugin1993}.
As a result, galaxies with 
a given oxygen abundance can have different N/O ratios and nitrogen abundances,  
%(and, consequently, different values of 
%the nitrogen abundance) 
depending on their star formation histories. 
We now check that expectation with our data. We use subsample C to have the most statistics. Fig.\ref{figure:znod} shows the oxygen and nitrogen abundances as a function of
redshift for galaxies with masses in the lower range 10$^{10.0}$M$_\sun$--10$^{10.3}$M$_\sun$ (upper left panel)  
and in the upper range 10$^{11.2}$M$_\sun$--10$^{11.5}$M$_\sun$ (upper right panel). 
The N/O abundance ratios of those 
galaxies as a function of redshift are shown in the lower panels.  
It can be seen that the galaxies in the two different mass ranges show 
different behaviors with redshift, reflecting 
 the difference in the level of star formation activity in them. 
%The level of star formation activity in galaxies 
That in the lower mass range  
%10$^{10.0}$M$_\sun$--10$^{10.3}$M$_\sun$ 
is high (those galaxies 
show an appreciable increase of oxygen abundance with decreasing redshift) 
while that in galaxies in the upper mass range is very low (those galaxies 
do not show a significant oxygen enrichment with decreasing redshift).  
Fig.\ref{figure:znod} shows that, in spite of the different behaviors of the oxygen abundances with redshift in  the two galaxy mass ranges, 
%galaxies with masses 10$^{11.2}$M$_\sun$--10$^{11.5}$M$_\sun$ and  
%in the galaxies with masses 10$^{10.0}$M$_\sun$--10$^{10.3}$M$_\sun$ show the 
%different general trends with redshift, 
the scatter in oxygen abundances at a given 
redshift is comparable in the two cases, the dispersion of the data points in the low mass range being only slightly larger than that in the high mass range. 
By contrast, the scatter in the N/O ratios and in the nitrogen 
abundances at a given redshift is considerably higher for galaxies with high star formation 
activity (those in the low mass range, left 
upper and lower panels in Fig.\ref{figure:znod}) than for galaxies with low star formation 
activity (those in the high mass range, right upper and lower 
panels in Fig.\ref{figure:znod}). 
At z=0.05, oxygen abundances in galaxies with masses 10$^{10.0}$M$_\sun$--10$^{10.3}$M$_\sun$ 
are similar to those in galaxies with masses 10$^{11.2}$M$_\sun$--10$^{11.5}$M$_\sun$. 
The maximum value of the N/O ratio in the two galaxy mass ranges is also similar. 
%galaxies with masses 
%10$^{10.0}$M$_\sun$--10$^{10.3}$M$_\sun$ is similar to the 
%value of N/O ratio in the galaxies with masses 10$^{11.2}$M$_\sun$--10$^{11.5}$M$_\sun$. 
However, many galaxies in the low mass range 
%with masses 
%10$^{10.0}$M$_\sun$--10$^{10.3}$M$_\sun$ 
possess low N/O ratios 
(Fig.\ref{figure:znod}, lower left panel). 
The fact that the N/O ratios 
of many galaxies with high star formation activity are  
significantly lower than those in galaxies with low star formation activity 
%do not show a low N/O ratio values, taken together  
does confirm our expectation that  
the larger scatter of nitrogen abundances around the z--M$_S$--N/H in comparison to 
that of oxygen abundances around the z--M$_S$--O/H is caused 
by the temporary decrease of the N/O ratio, due to star formation 
bursts of different amplitudes and/or ages, i.e. 
to different star formation histories in galaxies.

Thus, the consideration of the nitrogen abundance evolution with redshift 
and galaxy stellar mass has confirmed the general picture obtained 
from the oxygen abundance evolution analysis. 
Examination of both the z--M$_S$--N/H and 
the z--M$_S$--O/H relations has led to the following conclusions: \\
-- The galaxies of highest masses, those with masses $\ga$ 10$^{11.2}$M$_\sun$, 
have reached their high astration level 
more than 3 Gyr ago, so that stars in those galaxies have returned their 
nucleosynthesis products to the interstellar medium before $z$=0.25. \\
-- The galaxies with masses in the range from $\sim$ 10$^{11.0}$M$_\sun$  
to $\sim$ 10$^{11.2}$M$_\sun$ also form their stars before $z$=0.25, but later 
in comparison to the galaxies of highest masses. 
The intermediate-mass stars in those galaxies have not returned nitrogen to the 
interstellar medium before $z$=0.25 because they have not had enough 
time to evolve. \\
-- Significant star formation has occurred in galaxies  with masses lower than  
$\sim$ 10$^{11}$M$_\sun$ during the last 3 Gyr. 
Those galaxies have converted up to 20\% of their total mass into stars over 
this period. \\
-- Stars with lifetimes of a few Gyr contribute to the nitrogen production.

\subsection{Influence of the selection criteria on the redshift -- galaxy mass -- 
metallicity relations}

%We have discussed above that 
%our selecting galaxies by their strong emission lines is equivalent to imposing 
%a lower limit on their gas mass 
%fraction. 
%The mean $\mu$ in the most-evolved SDSS galaxies in our selected subsamples 
%was estimated to range from $\sim$ 5\% to $\sim$ 25\%, with a mean of $\sim$ 15\%
%It can be considered as evidence in favor of that the galaxies with gas mass 
%fraction less than $\sim$0.05 undergo the sufficiently strong star formation 
%bursts were rare (if ever). 
%Thus,  galaxies with a gas mass fraction less than $\sim$0.05 do not undergo sufficiently strong star 
%formation bursts to be included in our SDSS subsamples. These include galaxies with astration levels %as high as 0.9-0.95. 
%Therefore, one can expect that the criteria to  
%selection of ``emission-line galaxies'' do not influence dramatically 
%the redshift -- galaxy mass -- metallicity relations derived. 

We wish to examine here how our selection criteria affect the results obtained. 
We compare the results obtained for our SDSS subsamples with those obtained for a sample of star-forming SDSS galaxies selected with commonly used criteria.  
We use the diagnostic diagram proposed by \citet{baldwinetal1981} where 
the excitation properties of H\,{\sc ii} regions are studied by 
plotting the low-excitation [N\,{\sc ii}]$\lambda$6584/H$\alpha$ 
line ratio against the high-excitation [O\,{\sc iii}]$\lambda$5007/H$\beta$ 
line ratio.  The diagram can be used to separate  
different types of emission-line objects according to their main  
excitation mechanism. We have thus excluded from the total SDSS sample 
all objects above the solid line given by the equation
\begin{equation}
\log (\mbox{\rm [O\,{\sc iii}]$\lambda$5007/H$\beta$}) =
\frac{0.61}{\log (\mbox{\rm [N\,{\sc ii}]$\lambda$6584/H$\alpha$})-0.05} +1.3
\label{equation:kauff}   
\end{equation}
which separates objects with H\,{\sc ii} spectra from those 
containing an AGN \citep{kauffmannetal2003}.
We have extracted from the total sample a subsample of 95,046 star-forming 
galaxies, referred to hereafter as the "all star-forming galaxies" or SFG subsample. 

\begin{figure}
\begin{center}
\resizebox{0.80\hsize}{!}{\includegraphics[angle=0]{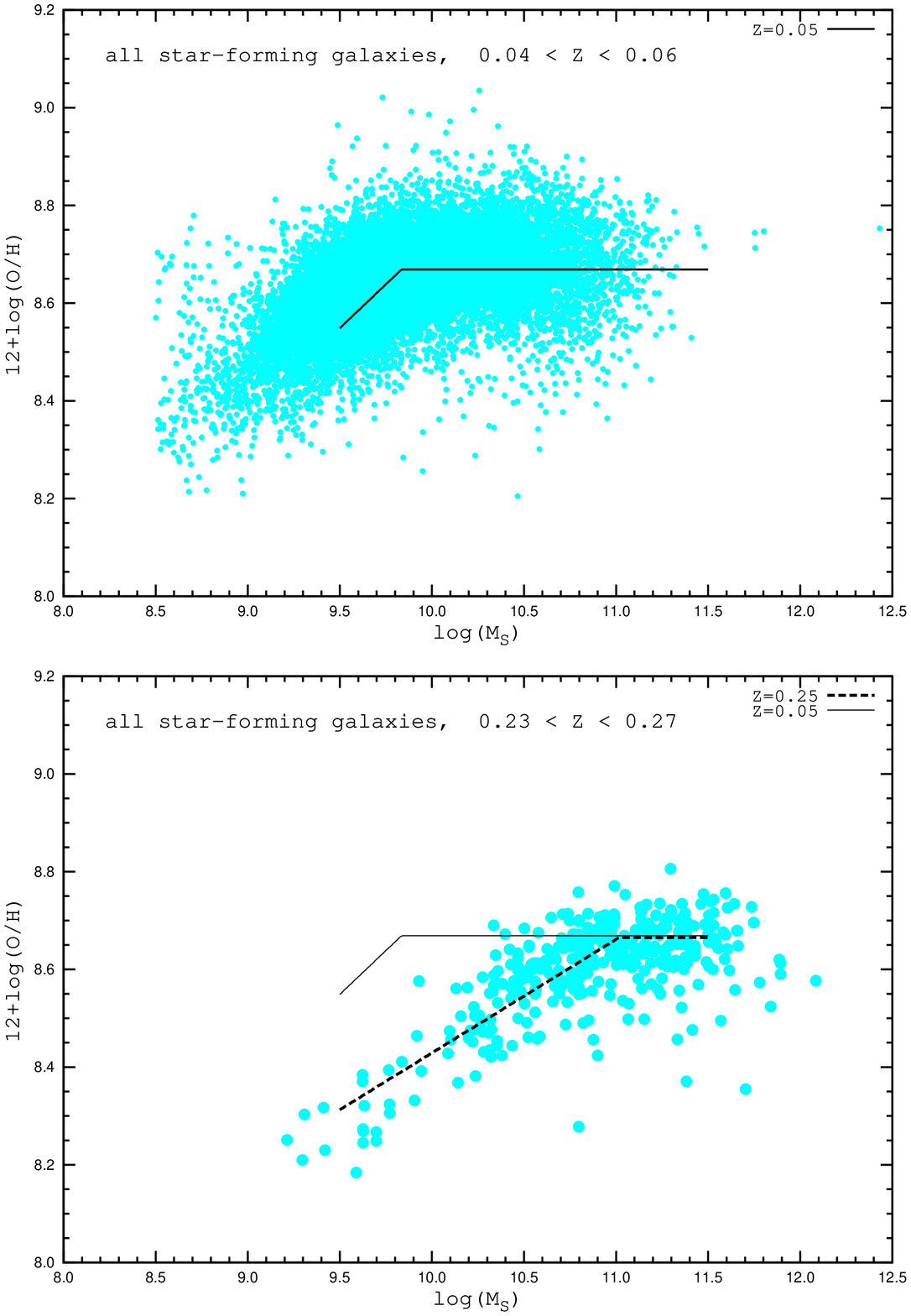}}
\caption{
The oxygen abundance -- galaxy stellar mass diagrams for the "all star-forming
galaxies" (SFG) subsample at $z$=0.05 (upper panel) and $z$=0.25 (lower panel). 
The z-M$_S$-O/H relation derived for subsample C (Eq.\ref{equation:yzmohc}) 
is shown by the solid line  for $z$=0.05, and by the dashed line for $z$=0.25.
}
\label{figure:zmohd}
\end{center}
\end{figure}

We compare the z--M$_S$--O/H and z--M$_S$--N/H diagrams for the SFG and C subsamples. 
The upper panel of Fig.\ref{figure:zmohd} shows the oxygen abundances 
of galaxies in the SFG subsample, with redshifts 
in the range 0.04 $<$ $z$ $<$ 0.06, as a function of galaxy stellar mass. 
The lower panel of Fig.\ref{figure:zmohd} shows the same diagram, but for 
galaxies in a higher redshift range, 0.23 $<$ $z$ $<$ 0.27.
The derived z--M$_S$--O/H relations for subsample C, superimposed on the data points of the SFG 
subsample,   
are shown in Fig.\ref{figure:zmohd} by the solid 
line for $z$=0.05, and by the dashed line for $z$=0.25.
Inspection of Fig.\ref{figure:zmohd} shows that the galaxies from 
the SFG subsample follow well the 
z--M$_S$--O/H relation derived for subsample C. 
Comparison of Fig.\ref{figure:zmohd} with Fig.\ref{figure:zmohc} shows however 
that the scatter of the points about the  
z--M$_S$--O/H relation is larger for the SFG subsample than for subsample C. 
This is evidence that our selection criteria effectively weeds out galaxies with less reliable 
measurements.

\begin{figure}
\begin{center}
\resizebox{0.80\hsize}{!}{\includegraphics[angle=0]{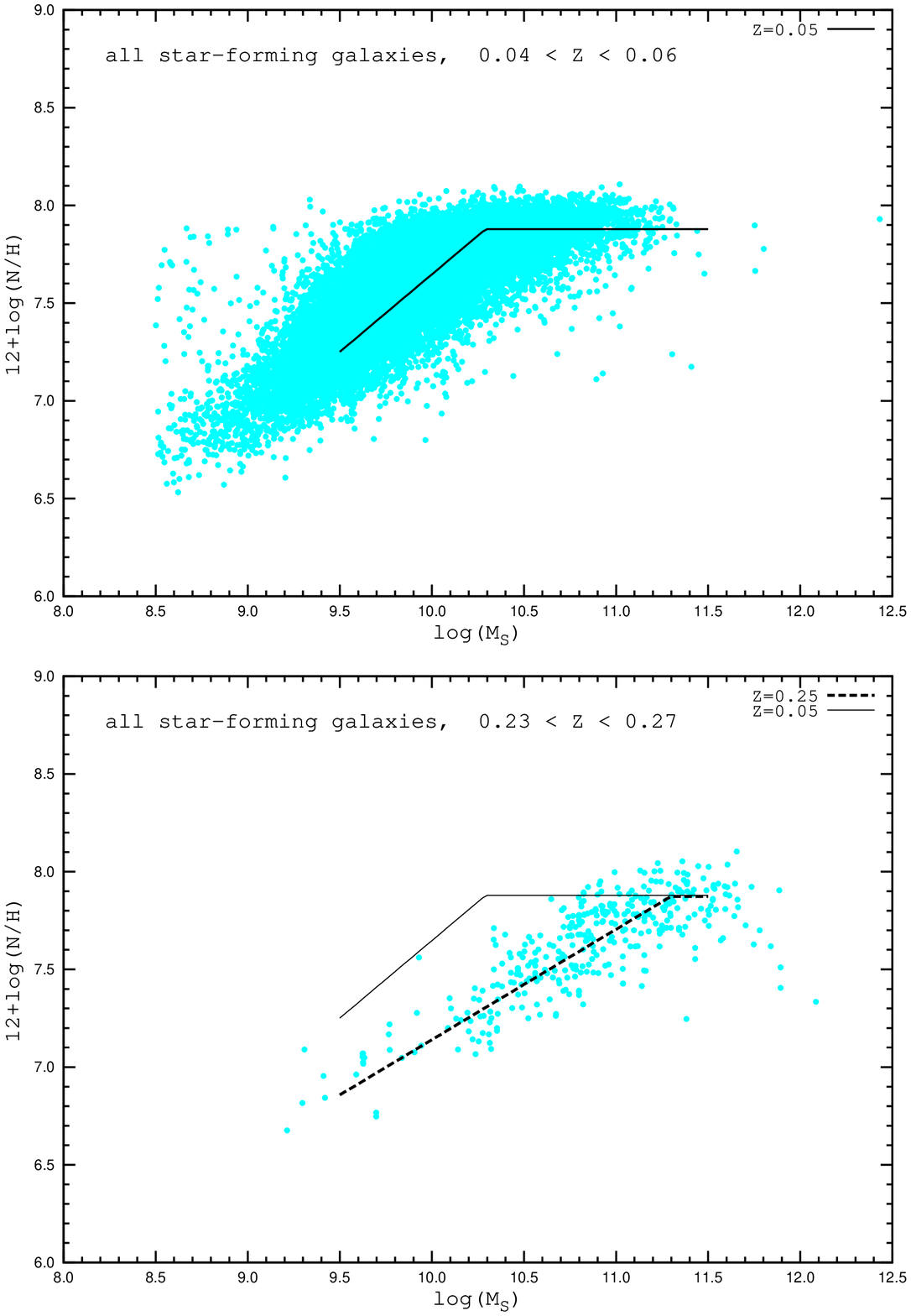}}
\caption{
The nitrogen abundance -- galaxy stellar mass diagrams for the "all star-forming 
galaxies" (SFG) subsample   
at $z$=0.05 (upper panel) and $z$=0.25 (lower panel). 
The z-M$_S$-N/H relation derived for subsample C (Eq.\ref{equation:yzmnhc}), 
is shown by the solid line  for $z$=0.05,  
and by the dashed line for $z$=0.25.
}
\label{figure:zmnhd}
\end{center}
\end{figure}

Fig.\ref{figure:zmnhd} shows the z--M$_S$--N/H diagram. 
As before, the z--M$_S$--N/H relations derived for subsample C are superimposed 
on the datapoints of the SFG subsample. 
Again, the galaxies from 
the SFG subsample follow well the 
z--M$_S$--N/H relation derived for subsample C. 
Thus, the results obtained with the commonly used method of selecting  star-forming galaxies 
do not differ appreciably from those obtained with our selection method. 
We emphasize that our approach presents however two important advantages.
First, it is not necessary to know a priori the precise   
location of the dividing line between H\,{\sc ii} regions 
and AGNs, a subject which is, as noted before, still controversial  
\citep{kewleyetal2001,kauffmannetal2003,stasinskaetal2006}.
Second, our approach allows to reject all unreliable measurements.
The counterpart of these advantages is that our approach requires a good sample of 
calibration datapoints -- a reasonably large sample of H\,{\sc ii} regions with 
precise spectrophotometric measurements, including weak auroral 
lines -- to establish reliable relations between the strong 
line fluxes and the physical characteristics of H\,{\sc ii} regions. 
Those relations should however be established in any case, as they are necessary for 
abundance determinations in star-forming galaxies, 
independently of the particular method used to select them.

\subsection{Comparison to previous work on the z--M$_S$--O/H relation}

The evolution of the mass-metallicity relation of galaxies  
with redshift has been considered previously by several groups, as 
described in the introduction. 
But, as discussed before, each group uses a different calibration to derive 
abundances which show as a result sometimes large discrepancies. 
So it is difficult to directly compare, or put on the same scale, 
the abundances derived by other groups and our own. Thus, we will not 
attempt such a comparison. Rather, we will limit ourselves 
to comparing the evolution in oxygen abundances with redshift 
which seems to be less sensitive to the adopted calibration. 
As for the nitrogen evolution with redshift, it has not been considered before. 
 
 We first summarize the results obtained by previous investigators.
\citet{lillyetal2003} have estimated the oxygen abundance in a sample of 
66 Canada--France Redshift Survey galaxies in the redshift range 0.47 $<$ $z$ $<$ 0.92, 
using the flux ratios of bright oxygen emission lines. They concluded 
that, at half the present age of the universe, the overall oxygen abundance 
of the galaxies in their sample, with luminosities ranging 
from M$_B$=--20 to M$_B$=--22 (or log(L$_B$/L$_{B\sun}$) 
$\sim$ 10.2 -- 11.0),  is only slightly lower than the oxygen abundance in similar luminous 
galaxies today. They found a variation $\Delta$(log(O/H)) = 0.08$\pm$0.06. 

\citet{savaglioetal2005} have investigated the mass--metallicity relation 
using galaxies from the Gemini Deep Deep Survey and the Canada--France Redshift Survey 
in the redshift range 0.4 $<$ $z$ $<$ 1.0. 
Their galaxies with  M$_S$ $>$ 10$^{10}$M$_{\sun}$  have oxygen abundances close to 
those in local galaxies of comparable mass, while the oxygen abundances in galaxies 
 with  M$_S$ $<$ 10$^{10}$M$_{\sun}$  are lower on average (with a large scatter) 
than those in galaxies of similar masses at the present epoch 
 (see their Fig.13).

\citet{cowiebarger2008} have studied the oxygen abundance evolution from 
$z$ = 0.9 to $z$ = 0.05 using a large sample of galaxies in the Great Observatories 
Origins Deep Survey--North (GOODS--N). They have found an evolution of 
the metallicity--mass relation corresponding to a decrease of 0.21$\pm$0.03 dex 
between the value at $z$ = 0.77 and the local value in the 
10$^{10}$--10$^{11}$M$_{\sun}$ range. They also found that star formation in the 
most massive galaxies ($>$10$^{11}$M$_{\sun}$) ceases at $z$$<$1.5 
because of gas starvation.

\citet{lamareilleetal2009} have derived the mass--metallicity relation of 
star-forming galaxies up to $z$ $\sim$ 0.9 using data from the VIMOS VLT Deep Survey.
They found that the galaxies of 10$^{10.2}$ solar masses show a larger 
oxygen enrichment ($\Delta$(log(O/H)) $\sim$ 0.28) from $z$ $\sim$ 0.77 
to $z$=0) than the galaxies of 10$^{9.4}$ solar masses  
($\Delta$(log(O/H)) $\sim$ 0.18). 

\citet{laralopezetal2009} have studied the oxygen abundance of relatively massive 
(log(M$_S$/M$_\sun$) $\ge$ 10.5) star-forming galaxies from SDSS/DR5 at different 
redshift intervals from 0.4 to 0.04. They found an oxygen enrichment 
$\Delta$(log(O/H)) $\sim$ 0.1 from redshift 0.4 to 0.

\citet{asarietal2009} have derived the mass--metallicity relation at different 
lookback times for SDSS galaxies using the stellar metallicities estimated with 
their spectral synthesis code. They have found that the more massive galaxies 
show very little evolution since a lookback time of 9 Gyr. 

Examination of all these studies shows good qualitative agreement between them, 
but with a rather large scatter in the estimated values of the oxygen 
enrichment. The results of these investigations can be summarized as followed: 
1) the most massive galaxies (those with masses $>$ 10$^{11}$M$_{\sun}$) do not show 
an appreciable enrichment in oxygen from $z$ $\sim$ 0.7 to $z$ = 0; 
2) in the 10$^{10}$--10$^{11}$M$_{\sun}$ mass range, an increase of the oxygen abundance 
$\Delta$(log(O/H)) $\sim$ 0.08--0.28 is observed in the redshift range from $z$ $\sim$  0.7 
to z $\sim$ 0.
Our results are also in good qualitative agreement with those previous results. We also see 
no change in oxygen abundance with redshift for galaxies with masses greater 
than 10$^{11}$M$_{\sun}$, in the redshift range from $\sim$0.25 to 0.
We estimate the mean increase of the oxygen abundance 
with redshift in the 10$^{10}$--10$^{11}$M$_{\sun}$ galaxy stellar mass 
range to be  
$\Delta$(log(O/H)) $\sim$ 0.11, with $\Delta$(log(O/H)) $\sim$ 0.23 at 
10$^{10}$M$_{\sun}$ and $\Delta$(log(O/H)) =0  
at 10$^{11}$M$_{\sun}$. This is in agreement with the upper range of  
values found in previous works, especially when we take into account 
the fact that we have 
considered a smaller redshift interval than previous investigators. 
%The slightly greater evolution of the oxygen abundance found here may be the result of our selection %criteria which favor strong ongoing 
%star formation. Conceivably, the star formation rate in our galaxies is higher than average for 
%galaxies of a given mass, which would result in a larger than average evolution in oxygen 
%abundances.  

\subsection{Nitrogen enrichment by intermediate mass stars: a discrepancy with current stellar evolution theory}

\citet{edmundspagel1978} have suggested  that observations of the N/O abundance 
ratio in galaxies can be understood if N is manufactured in stars of 
1--2.5 M$_\sun$. The N/O ratio of a galaxy then becomes an indicator of the time 
that has elapsed since the bulk of star formation occurred, or in other words 
of the nominal ``age'' of the galaxy.
\citet{pilyuginetal2003} have found that the N/O ratios in 
H\,{\sc ii} regions of galaxies of early morphological types are systematically 
higher than those in H\,{\sc ii} regions with the same O/H value in 
galaxies of late morphological types. Moreover, it is known that the 
star formation histories of galaxies of different morphological types 
differ, in the sense that the spiral galaxies of early morphological 
types have a significantly larger fraction of old stars than the galaxies of 
late morphological type \citep{sandage1986}. These two facts 
lead to the conclusion that the contribution  
to the nitrogen production by low- and intermediate-mass stars, with a long time delay 
of a few Gyr between their moment of birth and the time of release of 
their nucleosynthetic products in the interstellar medium, can be significant. 

One of the main conclusions of this study is that there has been a significant evolution in nitrogen abundances in star-forming galaxies  over the last 2--3 Gyr.  This appears not to agree with the predictions of current theoretical models of intermediate mass stars. Indeed, a time duration of 2--3 Gyr corresponds to the lifetime of stars in the 
1.5 -- 2 M$_\sun$ mass range. While stellar evolution models of intermediate mass stars have long predicted that they contribute significantly to nitrogen enrichment, it is stars of 3 -- 8  M$_\sun$ that are 
supposed to do the job, {\it not} stars in the 1.5 -- 2 M$_\sun$ mass range. Intriguingly, there are other types of observations that also suggest that stars in the 1.5 -- 2 M$_\sun$ mass range contribute to nitrogen enrichment.  
\citet{richermccall2008} have found that planetary nebulae in nearby galaxies (with 
or without ongoing star formation)  
often show large nitrogen enrichments. They have argued that 
these planetary nebulae are the descendants of relatively low mass progenitors, 
of approximately 1.5M$_\sun$ or less, i.e. that low- and intermediate-mass stars 
are a more important source of nitrogen than has been hitherto considered.
Thus, our conclusion that stars with lifetimes of a few Gyr contribute to the 
nitrogen production, derived from the consideration of the nitrogen abundance 
evolution with redshift in galaxies,  is in line with the conclusions of other types of 
investigations. While stellar evolution theory does not yet predict nitrogen production in
stars with masses of $\sim$ 1.5 -- 2 M$_{\sun}$, several lines of observations now do so.

\section{CONCLUSIONS}

The redshift evolution of oxygen and nitrogen abundances in emission-line 
SDSS galaxies has been studied. 

We have paid particular attention to the construction of our galaxy sample, 
using the MPA/JHU catalogs of line flux measurements and other derived 
physical properties for SDSS galaxies. We have devised a way to recognize 
and exclude from consideration
not only AGNs, but also star-forming galaxies with 
large errors in their line flux measurements.  
We have found that the requirement that nitrogen abundances, derived with 
different calibration relations based on different emission lines, 
agree, can be used as a reliable criterion to select star-forming 
galaxies with accurate line fluxes measurements. 

We have derived relations between nitrogen abundances and the abundance-sensitive  
N$_2$, N$_2$/R$_2$, and N$_2$/R$_3$ indexes. Those relations 
have been used to determine nitrogen abundances in 
the SDSS galaxies. The small dispersion among the various derived nitrogen 
abundances for a given galaxy is used as a criterion to select star-forming 
galaxies with accurate line fluxes measurements. 
The relations between the electron temperature t$_2$ and the  
N$_2$, N$_2$/R$_2$, and N$_2$/R$_3$ indexes have also been established. 
Those calibrations have been used to estimate the N/O ratio  
and derive the oxygen abundances O/H from the nitrogen abundances N/H.

Subsamples of star-forming SDSS galaxies have been extracted from the MPA/JHU catalogs, 
using the small nitrogen abundance dispersion criterion described above.   
The nitrogen and oxygen abundances are estimated for these galaxies. 
The evolution of the oxygen and nitrogen abundances with redshift and 
galaxy stellar mass of galaxy are investigated, that of nitrogen abundances for the first time.
We have obtained the following main results. \\
1) The galaxies of highest masses (those more massive than $\sim$ 10$^{11.2}$M$_\sun$) 
do not show an appreciable enrichment in both oxygen and nitrogen 
from $z$=0.25 to $z$=0.05. 
Those galaxies have reached their high astration level 
in such a distant past that their stars have returned their 
nucleosynthesis products to the interstellar medium before $z$=0.25. \\
2) The galaxies in the mass range from $\sim$ 10$^{11.0}$M$_\sun$  
to $\sim$ 10$^{11.2}$M$_\sun$ do not show an oxygen enrichment, but do show 
some enrichment in nitrogen.
Those galaxies also formed stars before $z$=0.25, but at a later epoch 
in comparison to the galaxies of highest masses. 
Their stars have not returned nitrogen to the 
interstellar medium before $z$=0.25 because they have not had enough 
time to evolve. \\
3) The galaxies with masses lower than $\sim$ 10$^{11}$M$_\sun$ show 
enrichment in both oxygen and nitrogen abundances over the redshift period from 
$z$=0.25 to $z$=0.05, i.e. during the last 3 Gyr. The oxygen enrichment increases 
with decreasing galaxy mass,  from 10$^{11}$M$_\sun$ to  M$_S$=10$^{9.9}$M$_\sun$.  
It reaches a value $\Delta$log(O/H) $\sim$ 0.25 at  M$_S$=10$^{9.9}$M$_\sun$ 
and slightly decreases with further decrease of galaxy mass. 
The nitrogen enrichment increases 
with decreasing galaxy mass, from $\sim$ 10$^{11}$M$_\sun$ to  $\sim$ 10$^{10.2}$M$_\sun$. 
It reaches a value $\Delta$log(N/H) $\sim$ 0.65 at $\sim$ 10$^{10.2}$M$_\sun$ 
and slightly decreases with further decrease of galaxy mass. 
Significant star formation has occurred in those galaxies during the last 3 Gyr. 
They have converted up to 20\% of their total mass to stars over 
this period. \\
4) Stars with lifetimes of 2--3 Gyr, i.e. in the 1.5 -- 2 M$_\sun$ mass range, contribute to the nitrogen production. This is not in agreement with current stellar evolutionary models of intermediate mass stars which predict that stars in the 3 -- 8 M$_\sun$ mass range do the job, not stars in the 1.5 -- 2 M$_\sun$ mass range. \\
5) The general picture of the oxygen abundance evolution  
with redshift and galaxy stellar mass obtained here and in previous work is confirmed 
and strengthened by consideration of the nitrogen abundance evolution.

\subsection*{Acknowledgments}

We are grateful to the referee for his/her constructive
comments.
L.S.P. thanks the hospitality of the Astronomy Department of the 
University of Virginia where part of this investigation was carried out. 
L.S.P. and I.A.Z. acknowledge the partial support of the Cosmomicrophysics-2 
project of the National Academy of Sciences of Ukraine.    
   The authors acknowledge the work of the SDSS team. 
Funding for the SDSS has been provided by the
Alfred P. Sloan Foundation, the Participating Institutions, the National
Aeronautics and Space Administration, the National Science Foundation, the
U.S. Department of Energy, the Japanese Monbukagakusho, and the Max Planck
Society. The SDSS Web site is http://www.sdss.org/.
The SDSS is managed by the Astrophysical Research Consortium (ARC) for
the Participating Institutions. The Participating Institutions are The
University of Chicago, Fermilab, the Institute for Advanced Study, the Japan
Participation Group, The Johns Hopkins University, the Korean Scientist Group,
Los Alamos National Laboratory, the Max-Planck-Institute for Astronomy (MPIA),
the Max-Planck-Institute for Astrophysics (MPA), New Mexico State University,
University of Pittsburgh, University of Portsmouth, Princeton University, the
United States Naval Observatory, and the University of Washington.

\clearpage

%++++++++++++++++++++++++++++++++++++++++++++++++++++++ Table  NH t2
\begin{deluxetable}{llcccc}
%\rotate
%\tabletypesize{\tiny}
%\tabletypesize{\scriptsize}
\tablewidth{0pc}
\tablecaption{Values of coefficients in Eq.(\ref{equation:nh}) for 12+log(N/H),  
and in Eq.(\ref{equation:t2}) for t$_2$\label{table:nht}} 
%\begin{center}
\tablehead{
Y value                                                  & 
X value                                                  & 
a$_0$                                                    & 
a$_1$                                                    & 
a$_2$                                                    & 
a$_3$                                                    }
\startdata 
12+log(N/H)   & log(N$_2$)        &    7.649   &   1.454  &   0.257   & --0.148         \\
12+log(N/H)   & log(N$_2$/R$_2$)  &    7.918   &   0.877  & --0.058   &   0.038         \\
12+log(N/H)   & log(N$_2$/R$_3$)  &    7.526   &   0.521  &   0.062   &   0.014         \\
              &                   &            &          &           &                 \\
t$_2$         & log(N$_2$)        &    0.778   & --0.354  &   0.001   &   0.044         \\
t$_2$         & log(N$_2$/R$_2$)  &    0.702   & --0.248  &   0.042   &   0.019         \\
t$_2$         & log(N$_2$/R$_3$)  &    0.802   & --0.172  &   0.003   &   0.011         \\
\enddata
%\tablenotetext{a}{in units of 12+log(X/H)}
%\tablenotetext{b}{in units of solar luminosity}
\end{deluxetable}

\newpage

% FIGURE 
%======================================= No 01    NN calibration

\newpage

% FIGURE 
%======================================= No 02     t2 calibration

\newpage

% FIGURE 
%======================================= No 03     SeaGull

\newpage

% FIGURE 
%======================================= No 04   lR3-lR2

\newpage

% FIGURE 
%======================================= No 05     O/H-N/O

\newpage

% FIGURE 
%======================================= No 06  Histograms of parameters

\newpage

% FIGURE 
%======================================= No 07     Z -  Ms

\newpage

% FIGURE 
%======================================= No 08   Z - Ms - OH   subsample B

\newpage

% FIGURE 
%======================================= No 09   Z - Ms - OH   subsample C

\newpage

% FIGURE 
%======================================= No 10   errors in  dOH  and  dNH

\newpage

% FIGURE 
%======================================= No 11    Z - Ms - NH   subsample B

\newpage

% FIGURE 
%======================================= No 12    Z - Ms - NH   subsample C

\newpage

% FIGURE 
%======================================= No 13   Z - NO        subsample D

\newpage

% FIGURE 
%======================================= No 14   Z - Ms - OH   subsample D

\newpage

% FIGURE 
%======================================= No 15    Z - Ms - NH   subsample D

\end{document}